\documentclass[twocolumn,tighten,times]{aastex631}

\usepackage{amsmath}
\usepackage[caption=false]{subfig}

\newcommand{\Ms}{sMBH}
\newcommand{\Mp}{pMBH}
\newcommand{\Mtot}{$M_{\mathrm{bin}}$}
\newcommand{\Mgd}{$M_{\mathrm{gd}}$}

\newcommand{\Msb}{$M_{\mathrm{sb}}$}

\newcommand{\vg}{$v_{\rm g}$}

\newcommand{\tevol}{$t_{\mathrm{DF}}$}

\newcommand{\q}{$q$}

\newcommand{\fg}{$f_{\mathrm{g}}$}
\newcommand{\Rbh}{$R_{\mathrm{b,h}}$}
\newcommand{\Rgh}{$R_{\mathrm{g,h}}$}

\received{July 20, 2021}
\revised{July 20, 2021}
\accepted{\today}
\submitjournal{ApJ}

\shorttitle{The Coalescence and \textit{LISA} Detection Rates of MBHBs from TNG50-3}
\shortauthors{Li et al.}

\begin{document}

\title{Massive Black Hole Binaries from the TNG50-3 Simulation: I. Coalescence and \textit{LISA} Detection Rates}


\author[0000-0002-0867-8946]{Kunyang Li}
\affiliation{School of Physics and Center for Relativistic
  Astrophysics, 837 State St NW, Georgia Institute of Technology,
  Atlanta, GA 30332, USA}   
\email{kli356@gatech.edu}

\author[0000-0002-7835-7814]{Tamara Bogdanovi{\'c}}
\affiliation{School of Physics and Center for Relativistic
  Astrophysics, 837 State St NW, Georgia Institute of Technology,
  Atlanta, GA 30332, USA}
\email{tamarab@gatech.edu}

\author[0000-0001-8128-6976]{David R. Ballantyne}
\affiliation{School of Physics and Center for Relativistic
  Astrophysics, 837 State St NW, Georgia Institute of Technology,
  Atlanta, GA 30332, USA}
\email{david.ballantyne@physics.gatech.edu}

\author[0000-0001-8128-6976]{Matteo Bonetti}
\affiliation{Dipartimento di Fisica G. Occhialini, Università di
  Milano-Bicocca, Piazza della Scienza 3, IT-20126 Milano, Italy}
\affiliation{INFN, Sezione di Milano-Bicocca, Piazza della Scienza 3, IT-20126 Milano, Italy}
\email{matteo.bonetti@unimib.it}


\begin{abstract}

We evaluate the cosmological coalescence and detection rates for
massive black hole (MBH) binaries targeted by the gravitational wave
observatory Laser Interferometer Space Antenna ({\it LISA}). Our
calculation starts with a population of gravitationally unbound MBH
pairs, drawn from the TNG50-3 cosmological simulation, and follows
their orbital evolution from kpc scales all the way to coalescence
using a semi-analytic model developed in our previous work. We find
that for a majority of MBH pairs that coalesce within a Hubble time
dynamical friction is the most important mechanism that determines
their coalescence rate. Our model predicts a MBH coalescence rate $\lesssim 0.45$~yr$^{-1}$ and a {\it LISA} detection rate $\lesssim 0.34$~yr$^{-1}$. Most {\it LISA} detections should originate from $10^{\rm 6} - 10^{\rm 6.8}\,M_{\rm \odot}$ MBHs in gas-rich galaxies at redshifts $1.6 \leq z \leq 2.4$, and have a characteristic signal to noise ratio SNR $\sim 100$. We however find a dramatic reduction in the coalescence and detection rates, as well as the average SNR, if the effects of radiative feedback from accreting MBHs are taken into account.  In this case, the MBH coalescence rate is reduced by $78\%$ (to $\lesssim 0.1$~yr$^{-1}$), and the \textit{LISA} detection rate is reduced by $94\%$ (to $0.02$~yr$^{-1}$), whereas the average SNR is $\sim 10$. We emphasize that our model provides a conservative estimate of the LISA detection rates, due to the limited MBH mass range in TNG50-3, consistent with other works in the literature that draw their MBH pairs from cosmological simulations.
\end{abstract}


\section{Introduction}
\label{sec:intro}
Massive black holes (MBHs) are known to reside at the centers of most
massive galaxies \citep{S1982, KR1995,M1998}, and the hierarchal
formation model of galaxy evolution predicts that massive galaxies are
built up through a series of mergers
\citep[e.g.,][]{White1978,White1991}. Thus, it is expected that
following a merger of two massive galaxies, the
individual MBHs find themselves orbiting in the gravitational
potential of the merger remnant galaxy. By interacting with
the stellar and gaseous background of the remnant galaxy, the
separation of some of these MBH pairs will shrink to the point that
they become strong gravitational wave (GW) emitters before eventually
coalescing into a single MBH. The GWs emitted by merging MBHs make
them important sources for the upcoming Laser Interferometer Space
Antenna \citep[\textit{LISA};][]{LISA2017}, which will survey the frequency
range of $100$~$\mu$Hz -- $100$~mHz.
%
%
The expected rate of \textit{LISA} detections is related
not only to the frequency of galaxy mergers, but also to the physical
processes within the remnant galaxy that bring the individual MBHs to
coalescence. It is therefore important
to understand the evolution of MBHs in post-merger galaxies
in order to anticipate the GW signals probed by the GW observatories. 

Once the MBHs are at separations of
$\sim 1$ kpc in the post-merger galaxy, dynamical friction (DF) by gas
and stars will dominate the decay of their orbits
\citep{BBR1980}. This process describes how gravitational deflection
of gas \citep{O1999,KK2007} or collisionless particles \citep[e.g.,
  stars and dark matter;][]{C1943,AM2012} leads to the formation of an
overdense wake trailing a moving MBH, exerting a gravitational pull onto the
MBH and sapping its orbital energy. The timescale for this stage of the
decay is determined by the properties of the two MBHs and their host
galaxy. The most important of these include the total mass, mass ratio
and initial orbits of the MBHs, and the distribution and kinematics of
the gas and stars in the host galaxy.  


Other
processes are expected to supplant DF in shrinking the orbit at
separations $\la 1$~pc. For
example, in stellar ``loss-cone" 
scattering \citep[e.g.,][]{Q1996, QH1997,Y2002} stars are scattered
away from the MBH binary (MBHB\footnote{The two MBHs are referred to
  as a binary when they are gravitationally bound, or as a pair prior
  to becoming bound. The separation at which the two MBHs become bound
  depends on the properties of the galaxy and the masses of the MBHs,
  but is most often at separations $\la 1$~pc.}), removing orbital
energy and hardening the binary. While the scatterings cause many
stars to be ejected, the loss-cone can be efficiently refilled
due to the triaxiality of most galaxy potentials \citep{Y2002, Khan2011, V2014,Gua2017}. 
In addition, if the post-merger galaxy is sufficiently gas rich, drag
on the binary by the surrounding circumbinary disk may also play an
important role for its orbital evolution at separations $\la 0.1$ pc
\citep[e.g.,][]{A2005, MP2005}. According to \citet{Haiman2009}
and \citet{Dotti2015}, MBHBs can sink efficiently towards the galactic
center through Type-II migration. When the separation falls below
$\sim 1000$ Schwarzschild radii GW emission begins to dominate the
orbital decay until coalescence \citep{KT1976, BBR1980}.   

In earlier work \citep[][hereafter LBB20a and LBB20b]{LBB20a,LBB20b}, we developed a
semi-analytic model to study the effects
of the galactic and orbital parameters on the inspiral time and
eccentricity evolution of MBH pairs due to gaseous and stellar DF at
kpc scales. The post-merger galaxies considered in these studies
spanned a wide range of properties --- from very gas rich to gas poor;
bulge-dominated to disk-dominated; rapidly spinning to slowly rotating
-- which allowed a detailed exploration of how the DF forces affected the evolution of a MBH pair over a wide range of conditions and scenarios. For example, we found that the separation of a MBH pair decays fastest in remnant galaxies with gas fractions $f_g < 0.2$ and a gas disk
 rotating near its circular speed. The evolution time is also shortened for MBH pairs with total mass $> 10^6 M_{\rm \odot}$ and mass ratios $q \geq 1/4$ moving in either circular prograde orbits or on very eccentric retrograde orbits.  Systems with these properties were more likely to have their MBH separations reach 1~pc in less than a Hubble time, increasing their chances of becoming a strong GW source.

Here, we present a new version of our semi-analytic model that continues the evolution of the MBHB below 1~pc by including the effects of the additional processes mentioned above (i.e., loss-cone scattering, viscous drag and GW emission). These additions allow us to self-consistently compute the evolution time-scale from kpc-scales to coalescence in a model post-merger galaxy.
An additional new component of this work is that instead of considering an arbitrary range of possible galaxy properties (as we did in LBB20a and LBB20b), we use the properties and redshifts of post-merger galaxies identified in one of the IllustrisTNG simulations \citep{Naiman2018} to characterize the model galaxies in which the MBHs evolve. This allows us to place the MBHB evolution in the cosmological context and to evaluate the dependence of the MBH coalescence and \textit{LISA} detection rates on the properties of merger galaxies and their MBH pairs.
  

Other groups also predicted the \textit{LISA} detection rate using cosmological simulations combined with semi-analytic models for the MBHB dynamics below the resolution limit. For example, \citet{Salcido2016} use results from the cosmological simulation suite EAGLE, and assume constant delay times between the galaxy merger and coalescence. They predict the \textit{eLISA} detection rate to be $\sim 2 \, {\rm yr^{\rm -1}}$, largely dominated by coalescences of seed black holes merging at redshifts between 1 and 2. \citet{Katz2019} also estimate the \textit{LISA} detection rate using the Illustris cosmological simulation, combined with a semi-analytic model presented by \citet{DA2017} and \citet{KBH2017}, used to evolve MBH orbital dynamics below $\sim $ kpc. They predict a \textit{LISA} detection rate of $\sim  0.5 - 1 \,{\rm yr^{-1}}$ for MBHs with masses larger than $10^{\rm 5} \,M_{\rm \odot}$. More recently, \citet{Chen2021, Chen2022} used the cosmological simulation ASTRID and found cosmological MBH merger rate in the range $0.3 - 2\, {\rm yr^{-1}}$. Similar approaches have been used by \citet{Degraf2020} and \citet{Cury2021}, who report that differences in black hole seeding and growth models lead to more than an order of magnitude difference in predicted MBH merger rates. 

Crucially, we also explore the effects of radiation feedback on the
coalescence and \textit{LISA} detection rates in this paper. Earlier
studies have shown that the radiation produced by each MBH can influence
the dynamics of the system \citep{KK2007, LBB20b}. For MBHs evolving in gas-rich backgrounds, the ionizing radiation emerging from the innermost parts of their accretion
flows can affect the gaseous DF wake and render gas DF
inefficient for a range of physical scenarios. MBHs in this regime
tend to experience a positive net force, speeding them up,
contrary to the expectations for gaseous DF without radiative feedback
\citep{PB2017,G2020,T2020}. As showed by LBB20b, negative gaseous DF
can lengthen the inspiral time of MBHs and significantly reduce the
chance of forming close MBH pairs, particularly for lower mass MBHBs,
an important source class for the \textit{LISA} observatory.


This paper is organized as follows. In Sect.~\ref{sec:methods} we
describe the main features of the model used to evolve the MBHs from
kpc-scales to coalescence, as well as the IllustrisTNG sample of
post-merger galaxies. Section~\ref{sec:time} presents the distribution
of evolution times from the suite of models, including the contributions spent in each phase of the orbital decay. Section~\ref{sec:merger_rate} shows how the predicted MBHB coalescence rates and fractions are impacted by the different galactic properties. Section~\ref{sec:property_lisa} shows the predicted \textit{LISA} detection rates and the properties of systems that may be detected by \textit{LISA}. In Section~\ref{sec:RF_lisa} we discuss the effect of radiation feedback on the MBH merger and \textit{LISA} detection rates. 
Finally, we discuss the implications of our findings in
Sect.~\ref{sect:discuss} and conclude in
Sect.~\ref{sec:concl}. In this work, we assume an updated cosmology consistent with that used in the TNG simulation ($\Omega_{\rm \Lambda,\,0}=0.6911$, $\Omega_{\rm m,\, 0}= 0.3089$, $\Omega_{\rm b, \, 0}= 0.0486$, $h=0.6774$), and $t_{\rm Hubble}=14.4$ billion yrs.

\section{Methods}
\label{sec:methods}
In this section, we first review how we parameterize the structure
of post-merger galaxies (see LBB20a for full details), and then
describe how we identify such galaxies in IllustrisTNG and convert
them into our paramertized form. Section~\ref{sub:evo} details the
evolution calculations to model the inspiral of a secondary MBH from
kpc-scales to coalescence.

\subsection{Model of the Remnant Galaxy}
\label{sub:galaxymodel}
We assume a galaxy merger produces a single remnant, with a stellar
bulge and gas disk\footnote{We omit the stellar disk since its impact
on the orbital evolution from DF is negligible (LBB20a).}, which
includes the MBH pair. The half-mass radii of the bulge and disk are
\Rbh\ and \Rgh, respectively. The primary MBH (\Mp; with mass $M_1$) is fixed at the center of the galaxy. The non-rotating bulge has a mass \Msb\ and
follows a coreless powerlaw density profile \citep[e.g.,][]{BT1987}
which cuts off at $2\times R_{\mathrm{b,h}}$, with the scale parameters proportional to $\log (M_1/10^5 M_{\odot})$~kpc. We consider the orbital
evolution of a bare, secondary MBH (\Ms) with mass $M_2 < M_1$, which is orbiting in the plane of
the gas disk. The total mass of the MBH pair is $M_{\mathrm{bin}}=M_1+M_2$ and the mass ratio is $q=M_2/M_1$.

The gas disk follows an exponential profile with a scale radius of $2
\times (M_1/10^5 M_{\odot})$~kpc \citep[e.g.,][]{BT1987}. As a
result, models with larger \Mtot\ have gas densities that decrease
more slowly with radius, impacting the orbital decay from gaseous
DF. The gas densities in the disk are determined by the overall gas fraction
($f_{\mathrm{g}}=M_{\mathrm{gd}}/(M_{\mathrm{gd}}+M_{\rm sb})$), where
\Mgd\ is the mass of the gas disk (defined as the mass within $2\times$\Rgh). The disk rotates with velocity $v_{\rm g}(r)$, defined in units
of the local circular velocity $v_{\rm c}(r)$.  In our nomenclature $v_{\mathrm{g}} > 0$ if the galaxy disk and \Ms\ are corotating and $v_{\mathrm{g}} < 0$ if they are counterrotating.  Therefore, our model for a
merger remnant galaxy containing a pair of MBHs is defined by seven
parameters: \Mtot\ and $q$ for the MBHs, \Msb\ and \Rbh\ for the stellar
bulge, and \Mgd, \Rgh\ and \vg\ for the gas disk. As described
below, we determine the values of these parameters from the properties of merger remnant galaxies
in one of the IllustrisTNG simulations.

\subsection{Massive Black Hole Mergers in the TNG50 Simulation}
\label{sub:TNG}
The IllustrisTNG suite includes 18 simulations in total that differ in
the physical size of the computational domain, the mass resolution,
and the complexity of the included physics \citep{Naiman2018, Nelson2018,
  Marinacci2018, Pill2018,Spring2018}. There are three physical
simulation volumes available: $(50\,{\rm cMpc})^3$, $(100\,{\rm cMpc})^3$, and $(300\,{\rm cMpc})^3$, which are referred to as TNG50, TNG100, and TNG300, respectively. The mass
resolution of TNG50 is a few hundred times higher than that of the
TNG300 simulation \citep{Nelson2019, TNG50_a, TNG50_b} and provides the most detailed look at
the structural properties of galaxies. Therefore, we use the TNG50
simulation to identify and characterize post-merger galaxies.




TNG50 itself consists of a series of lower-resolution realizations of the same volume. We specifically use data from TNG50-3 as the $z=0$
gravitational softening of the collisionless component (i.e., stars
and dark matter) is $1.15$~kpc which is consistent with the initial
separation of the MBHs in the orbital decay calculation (LBB20a). The
simulation assumes a heavy-seed MBH formation model
\citep{Loeb1994,Begelman2006, Latif2013, Habo2016, Ar2018, Dunn2018}
with a seed mass of $\sim 10^{\rm 6} M_{\rm \odot}$ \citep{TNG50_a}, which sets a firm lower-limit to the mass of the
MBHs in the simulation.


The parameters of the MBH pair in a merging galaxy are
extracted from the \textit{blackhole mergers and details} supplementary data
catalog of TNG50-3\footnote{See
\url{https://www.tng-project.org/data/docs/specifications }.}
generated by the Illustris Black Holes Post-Processing Module \citep{Ble2016, KBH2017}, with 2165 `mergers' in total. These `mergers'
correspond to MBH pairs that reach the separation of the
gravitational softening of the collisionless component ($\approx
1$~kpc). For each `merger' we extract the black hole ID of the more
massive MBH (\Mp ), the masses of each MBH
($M_1$, $M_2$), and the mass ratio ($q$). The redshifts ($z$) of each `merger'
are determined from the snapshot ID and is converted into cosmic time
in Gyr by $t_0=t_{\rm Hubble}/(1+z)^{1.5}$~Gyr. This determines the starting time for the
evolution calculations described in Sect.~\ref{sub:evo}.



The final galaxy structure data needed for the model are the bulge
mass and half-mass radius (\Msb, \Rbh), the gas disk mass and
half-mass radius (\Mgd, \Rgh) and the disk rotation speed \vg (see the previous subsection). The
size and mass properties are determined by
cross-matching the merger information with the subhalo catalog using
the black hole ID of the \Mp\ to identify the post-merger galaxy.
The masses are assigned to the appropriate values listed in the
\textit{SubhaloMassInRadType} catalog entry for each
galaxy\footnote{This catalog provides the stellar and gas masses within $2\times R_{\mathrm{b,h}}$ of each TNG50-3 galaxy, yet the radius of the gas disk in our model galaxies is defined as $2\times R_{\mathrm{g,h}}$. As $R_{\mathrm{g,h}}$ is generally larger than $R_{\mathrm{b,h}}$, the gas densities of our model will, on average, be underestimated. The lower gas density will decrease the efficiency of gaseous DF and may extend the evolution time of the MBH pairs.}. Similarly, the two radii are assigned to the corresponding
values from the \textit{SubhaloHalfmassRadType} catalog entry for
each galaxy after converting from comoving to physical units. As the TNG50-3 catalog do not record the rotation of the post-merger galaxies, we randomly assign a value in the range of $v_g = [0.7,0.9]$
as the gas disk rotational speed \citep{Rog1971, Allen1973, Cas1983,
  Begeman1991, Blok2008, SINS2009, Lindberg2014, TNG50_K}.

Out of 2165 total 'merger' events recorded in the simulations, there are 45 `mergers' in TNG50-3 with $M_{\rm gd}=0$, and 123 `mergers' with both $M_{\rm gd}=0$ and $M_{\rm sb}=0$ which are likely misidentified subhalos \citep{Ble2016, KBH2017, Katz2019}. We omit these 123 `empty' systems from our analysis, but we do simulate the orbital evolution of the 45 `mergers' with only stellar components and find they have no influence on the overall coalescence rate or {\it LISA} detection rate since the evolution times of the sMBHs in these models are longer than the Hubble time.


At this point, we have characterized 2042 models of post-merger galaxies from
the TNG50-3 data, each with a specific MBH pair with known masses. Each
galaxy is also associated with a redshift specifying when the MBH pair
is at a separation of $\approx 1$~kpc. As all these galaxy mergers
are extracted from a known cosmological volume, we will be able to
compute the rates of MBH mergers from this starting
dataset. We are now prepared to dynamically evolve the MBH pair in
each galaxy model to determine the times to coalescence.

\subsection{Dynamical Evolution of the MBH pairs}
\label{sub:evo}
The orbital evolution of the \Ms\ due to gaseous and stellar DF is
followed from a separation of $\approx 1$~kpc until the influence
radius of the MBHB ($R_{\rm inf}$), at which point the orbital decay
is dominated by loss-cone (LC) scattering, viscous drag (VD) from a
circumbinary disk, and GW emission. The calculation ends when the
separation is smaller than the innermost stable circular orbits of the two MBHs ($6G M_{\rm bin}/ c^{2}$).

The orbit of the \Ms\ is not closed, but we use the farthest and
closest approaches of the MBH to estimate
the eccentricity $e$ throughout the DF calculation. The procedure to
initialize the orbit is described by LBB20a, and can provide a range of
initial eccentricities $e_i$. In addition, the orbit of the \Ms\ can
either be prograde or retrograde with respect to the galaxy rotation. To consider the effects of different
orbital geometries, we run the evolution calculation four times for
each of the 2042 galaxy models: twice with $e_i < 0.2$ and twice with
$0.8 \le e_i \le 0.9$. In both cases, we consider a prograde and a
retrograde orbit. Our final model suite therefore contains 8168
individual orbital evolutions.

\subsubsection{Dynamical Friction ${\rm (\sim 1 kpc - {\it R}_{inf})}$}
\label{subsub:DF}
The calculation of the orbital decay due to stellar and gaseous DF is
described in detail by LBB20a,b. The stellar DF force exerted by the bulge is calculated using Eqs.~(5)-(7) in LBB20a, following the work of \citet{AM2012}. The velocity distribution of stars in the
bulge is assumed to be Maxwellian (see LBB20b, Eq. 2). Since gaseous
DF depends on the Mach number of the moving body \citep[e.g.,][]{KK2007}, the sound speed of the gas disk must be defined. The
temperature profile of the disk is taken to be $10^4$\,K above the
minimum temperature required by the Toomre stability criterion
\citep{T1964}. The gaseous DF force on the \Ms\ is then computed using
Eqs.~(10)--(12) of LBB20a, which results in the gaseous DF force that
is strongest when the velocity difference between the \Ms\ and gas
disk is close to the sound speed, $c_s$ \citep{O1999,KK2007}. We define the inspiral time of sMBHs from $\sim 1$ kpc to ${R_{\rm inf}}$ as \tevol.


\subsubsection{Loss-Cone Scattering}
\label{subsub:LC}
When the MBH separation reaches $R_{\rm inf}$ and the mass enclosed in
the orbit is twice the binary mass\footnote{The mass enclosed in the orbit is calculated by integrating the stellar density profile (Eq.~(1) of LBB20a) and gas density profile (Eq.~(3) of LBB20a).}, LC scattering
dominates over DF in removing orbital energy \citep{Sesana2006}. The
hardening of MBHB orbits by LC scattering can be approximately described by
\begin{equation}
\label{eq:LC1}
\left(\frac{df_{\rm orb}}{dt}\right )_{\mathrm{LC}}=\frac{3G^{4/3}}{2(2\pi)^{2/3}}\frac{H \rho_{\rm i}}{\sigma}M_{\rm bin}^{1/3}f_{\rm orb}^{1/3}
\end{equation}
and
\begin{equation}
\label{eq:LC2}
\left(\frac{de}{dt}\right )_{\mathrm{LC}}=\frac{G^{4/3}}{(2\pi)^{2/3}}\frac{HK \rho_{\rm i}}{\sigma}M_{\rm bin}^{1/3}f_{\rm orb}^{-2/3},
\end{equation}
where $f_{\rm orb}$ is the MBHB orbital frequency, $\sigma$ is the stellar 
velocity dispersion of the host galaxy, $\rho_{\rm i}$ is the stellar
density at $R_{\mathrm{inf}}$, and $H$ and $K$ are numerical factors from three-body scattering experiments \citep{Q1996, Sesana2006}.

Following \citet{Bonetti2019}, the stellar spheroid density profile
at $r<1$~pc can be described as 
\begin{equation}
\label{eq:rho_star}
\rho (r) = \frac{(3-\gamma)M_{\rm sb}}{4\pi}\frac{r_{\rm 0}}{r^{\rm \gamma}(r+r_{\rm 0})^{\rm 4-\gamma}}\,,
\end{equation}
where $\gamma = 1.8$, $r_{\rm 0} \approx 1.3R_{\rm eff}[2^{\rm
    1/(3-\gamma)}-1]$ \citep{Dehnen1993} and
\begin{eqnarray*}
  \label{eq:reff}
\log \left (\frac{R_{\rm eff}}{\rm kpc} \right ) =
\begin{cases}
  -5.54+0.56\log \left ( \frac{M_{\rm sb}}{{\rm M}_{\rm \odot}}\right )&  M_{\rm sb}>10^{\rm 10.3} {\rm M}_{\rm \odot}, \\
  -1.21+0.14\log \left ( \frac{M_{\rm sb}}{{\rm M}_{\rm \odot}}\right ) & M_{\rm sb} \leq 10^{\rm 10.3} {\rm M}_{\rm \odot}.
\end{cases}
\end{eqnarray*}
\begin{equation}
\end{equation}
The velocity dispersion of the stars is calculated from the
virial theorem $\sigma \sim 0.4(GM_{\rm sb}/r_{\rm 0})^{1/2}$ \citep{Baes2002}.

\subsubsection{Viscous Drag in a Circumbinary disk}
\label{subsub:VD}
When the sMBH enters any gaseous circumbinary disk, VD\footnote{Throughout the manuscript we refer to this evolution mechanism as “viscous drag”, for simplicity. Note however that angular momentum transport at the inner edge of the circumbinary disk is mostly driven by gravitational torques from the binary and not the viscous tourques.} may significantly contribute or even 
dominate the evolution of the binary. 
\citet{Haiman2009} described how the orbit of a MBHB embedded in a
circumbinary \citet{SS1973} $\alpha$-disk evolves due to VD, and how
this evolution depends on the different physical conditions within the disk. 
Here, we consider the regime that leads to the most efficient evolution of the binary:
a disk that is everywhere supported by gas pressure, with opacity
defined by electron scattering, and a radial extent of $1$~pc from the \Mp\footnote{This is a fiducial choice as the radial extent of MBHB circumbinary disks is not known. To mitigate the uncertainty we tested the impact of this parameter and found that the calculated MBH coalescence and detection rates change by less than 10\% for a 10 times smaller circumbinary disk.}. In this case the rate of evolution of orbital frequency in
the circumbinary disk due to VD is 

\begin{equation}
\label{eq:gas_drag}
\left(\frac{d f_{\rm orb}}{d t}\right)_{\mathrm{VD}} =\frac{\sqrt{G M_{\rm bin}} r^{\rm -3/2}}{(2.5\times 10^{\rm 5}\; {\rm yr}) \; M_{\rm 7}^{\rm 6/5}r_{\rm 3}^{\rm 7/5}},
\end{equation}
%
where $f_{\rm orb}$ is the orbital frequency, $r_{\rm 3}$ is the orbital semi-major axis in units of $10^{\rm 3}$ Schwarzschild radii, $M_{\rm 7}=\ M_{\rm bin}/10^{\rm 7} M_{\rm \odot}$, and $q_{\rm s}=4q/(1+q)^{\rm 2}$ is the symmetric mass ratio \citep{Haiman2009}. Note that this prescription implies that the MBHB orbit always shrinks under the influence of VD and does not account for a possibility of orbital expansion, investigated in some more recent simulations \citep{Tang2017, Miranda2017, Mu2019, Moody2019, Tiede2020, BB2021}.

The eccentricity evolution due to VD can be complex and cannot be trivially reduced to a prescription for a single dominant regime. \citet{Roedig2011} shows that if the incoming eccentricity
of the MBHB on a prograde orbit is $>0.04$ then there is a limiting
eccentricity in the range $(0.6,0.8)$ that the binary reaches during its
interaction with the circumbinary disk. Thus, if one of our model
MBHBs has a prograde orbit with an eccentricity larger than $0.04$
while VD dominates the evolution, we then randomly assign the eccentricity 
between $0.6$ and $0.8$ after one viscous timescale (measured at
the separation where VD begins to dominate the evolution). If however
the eccentricity of the orbit is less than $0.04$ when VD takes over
the orbital decay, the eccentricity remains fixed until GW emission
takes over the orbital evolution.  

For MBHBs in retrograde orbits, there are three possibilities for the
eccentricity evolution that depend on the value of the eccentricity
when the MBH reaches this stage \citep{Roedig2014}. If the \Ms\ is in a
near circular orbit (i.e. $e<0.04$), then its eccentricity
will not change due to VD. However, if $0.04 \leq e < 0.8$, the
eccentricity then increases as $\approx 0.09e-0.0034$ per orbit. Finally,
if $e \geq 0.8$, a disk-binary interaction causes the binary to leave the disk plane,
tilt, and converge to a prograde orbit with limiting eccentricity in
the range of $[0.6,0.8]$. The timescale for this transition corresponds
$\sim 10$ viscous timescales to reach the final steady state due to
the reversal of the orbital direction from retrograde to prograde
\citep{Roedig2014}. More recent works find qualitatively similar results: that circular binaries remain circular and that eccentric binaries tend to evolve toward a threshold eccentricity, which exact value depends on the thermodynamic properties of the disk and was found to be close to 0.4 by \citet{Dora2021} and \citet{Zrake2021}.

\begin{figure*}[t!]
    \centering
    \includegraphics[width=0.4\textwidth]{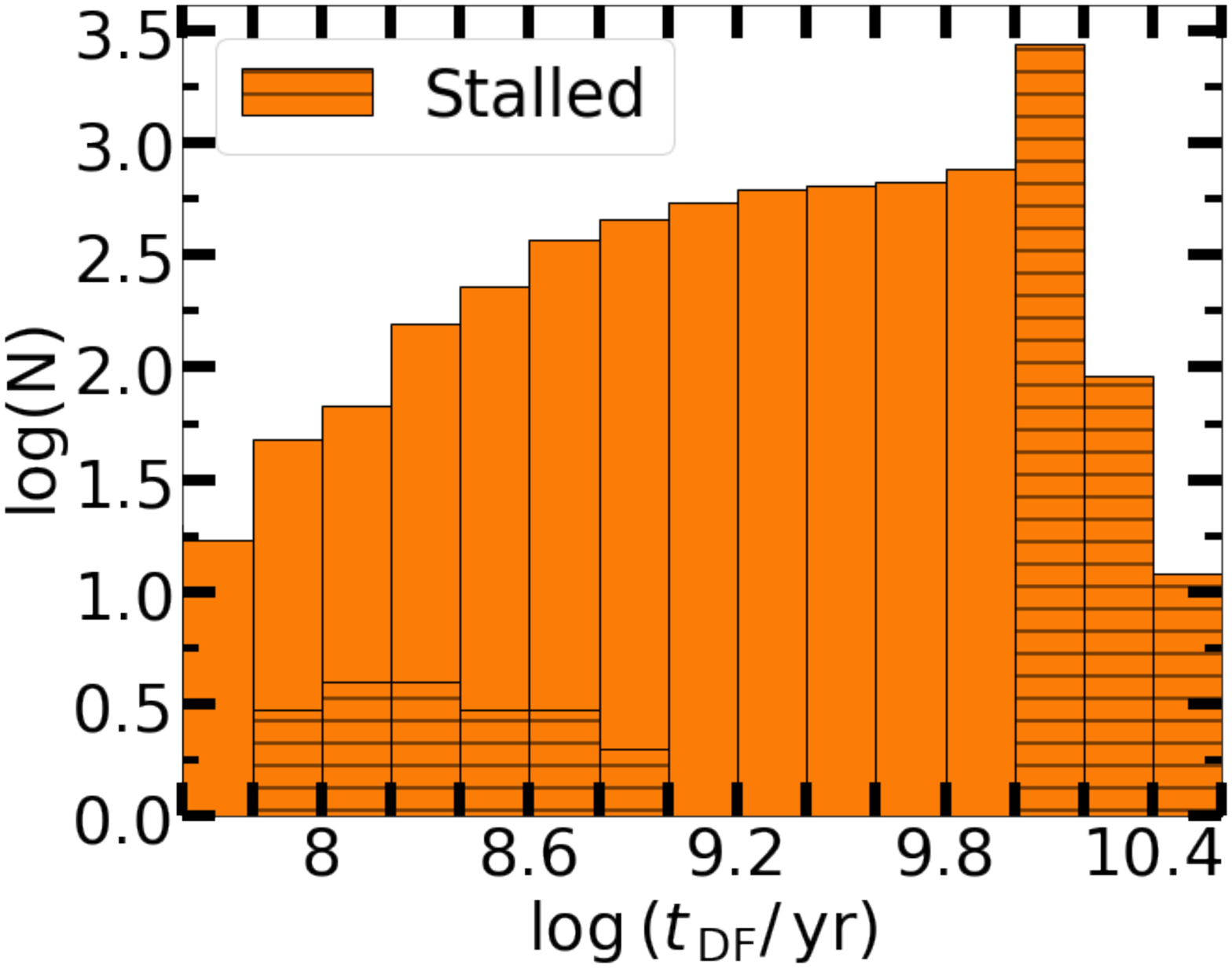}
    \includegraphics[width=0.4\textwidth]{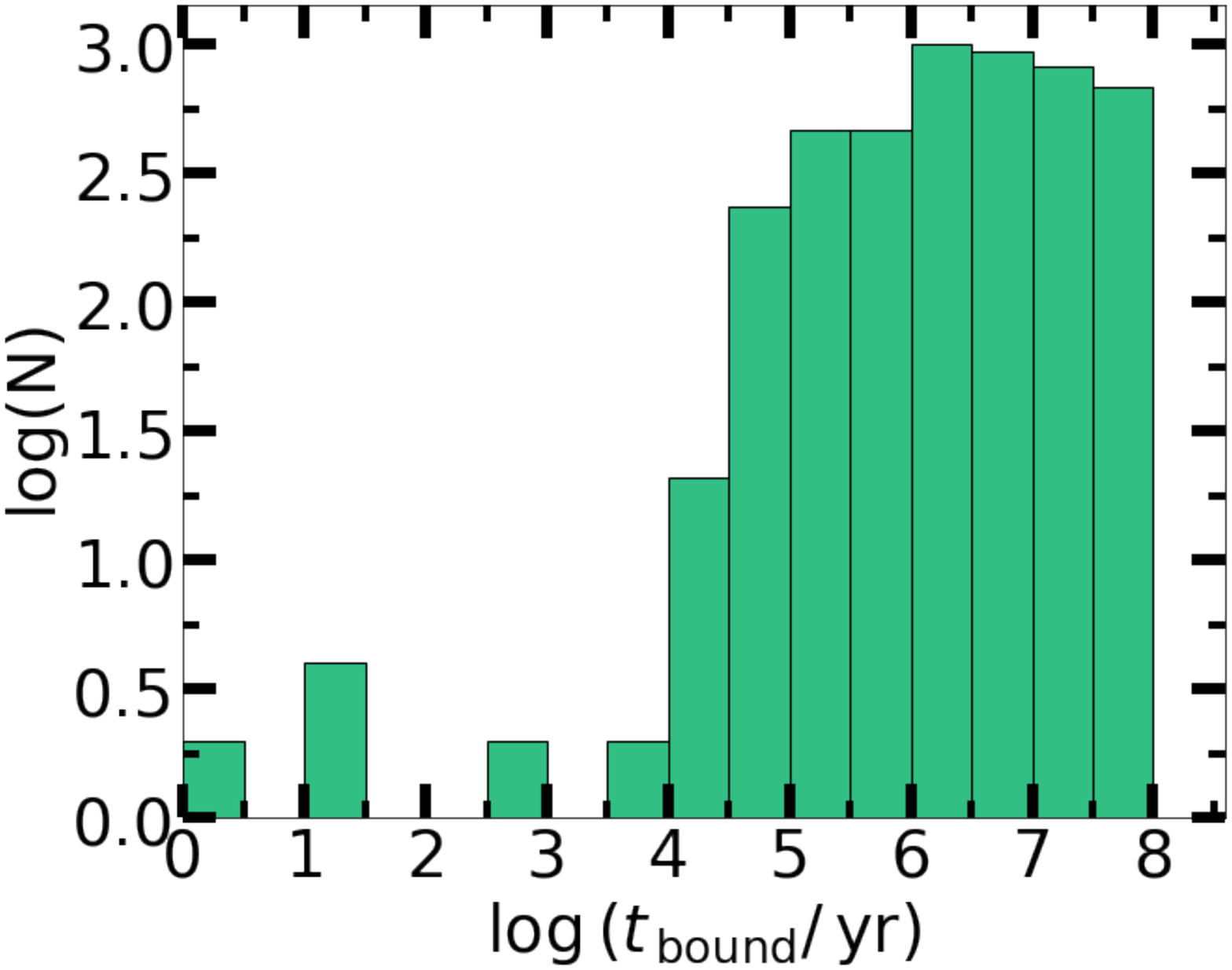}
    \includegraphics[width=0.4\textwidth]{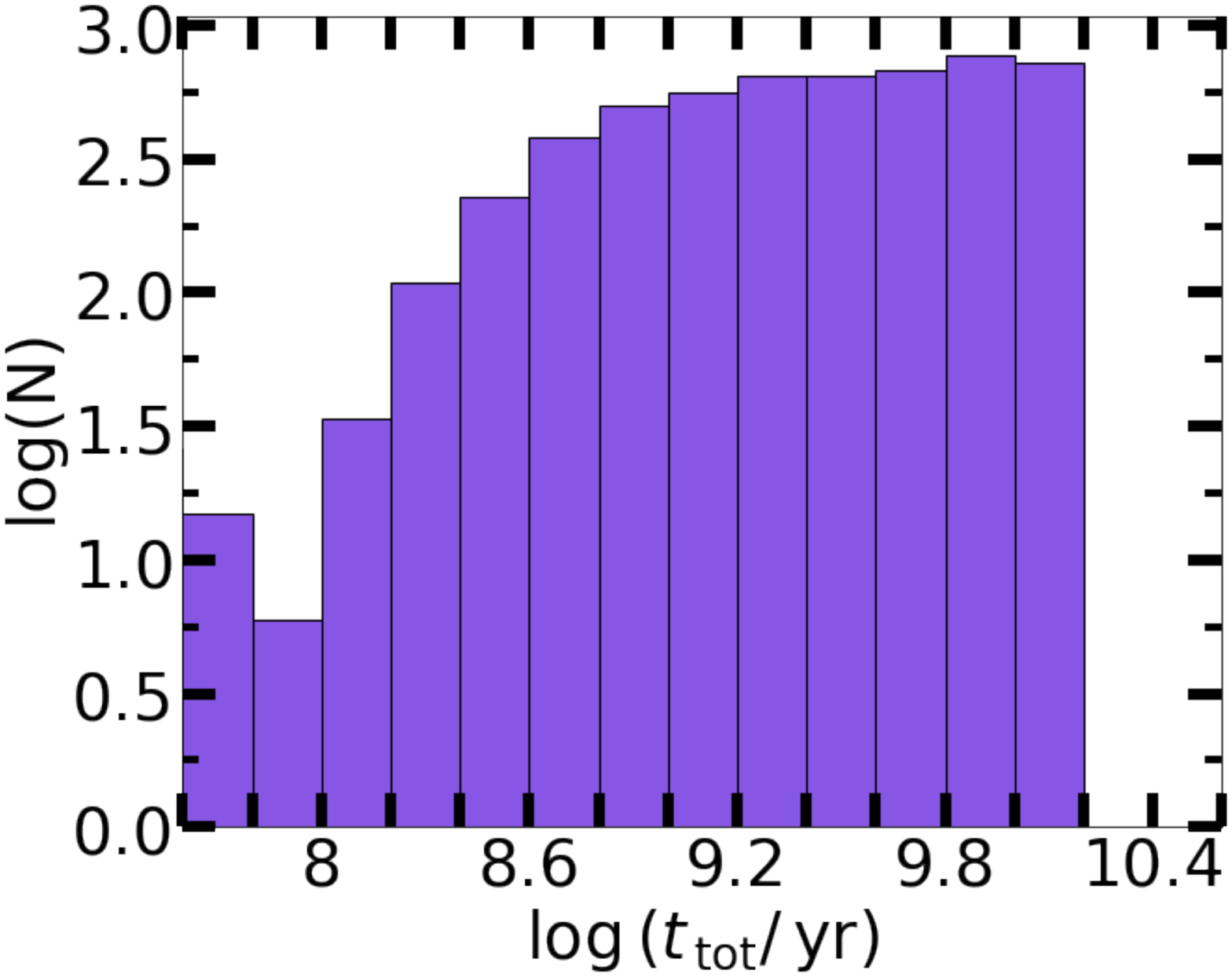}
    \includegraphics[width=0.4\textwidth]{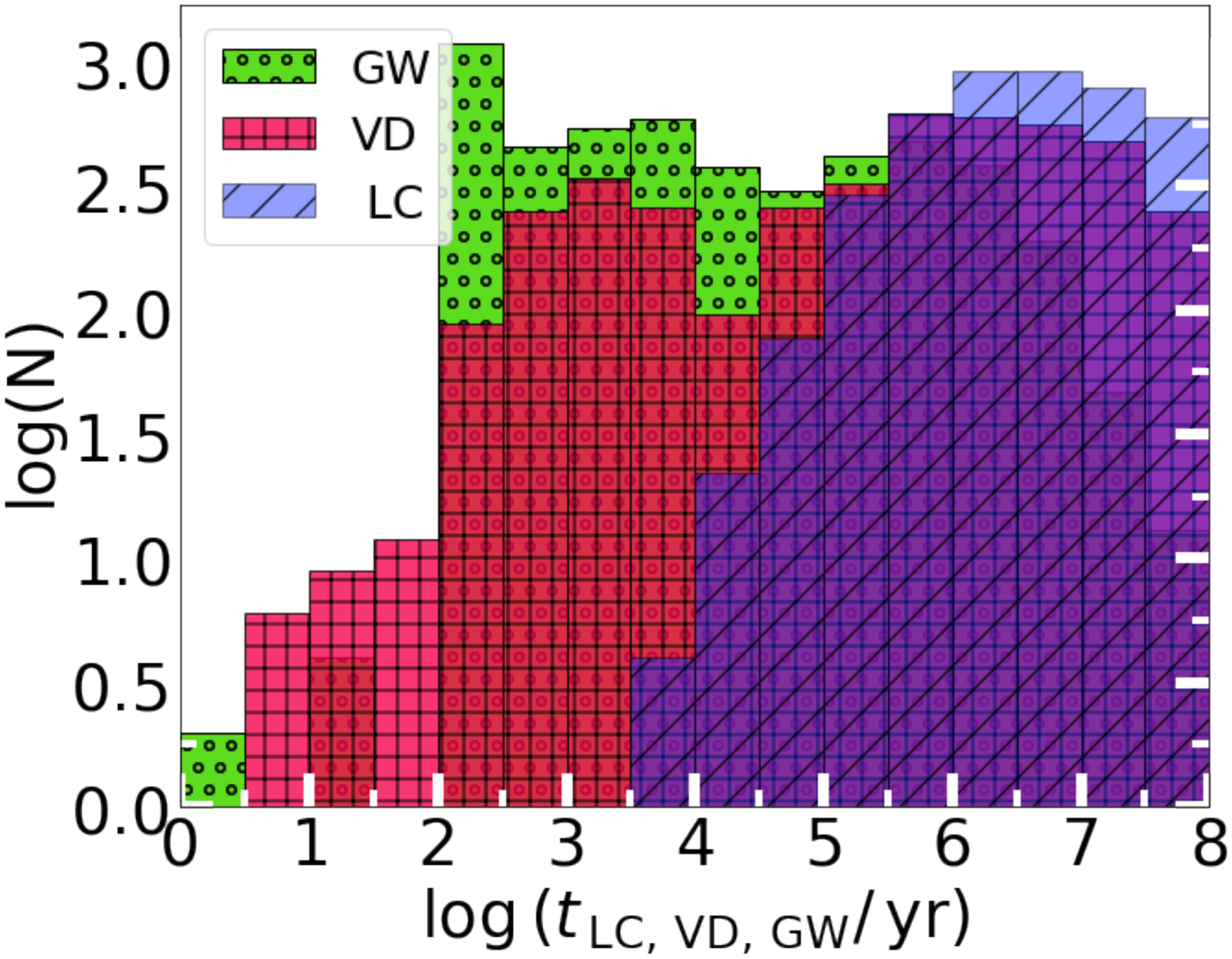}
\caption{Orbital evolution times of the MBH pairs
  in the model suite of 8168 orbital configurations. {\it Top left:}
  The distribution for the DF dominated stage, when the
  separation decays from $\sim 1$~kpc to
  $R_{\mathrm{inf}}$ ($t_{\mathrm{DF}}$). The hatched regions mark 
  models that did not coalesce in a Hubble time due to stalling
  in the DF phase (the right-hand side of the histogram), or within
  $R_{\mathrm{inf}}$ (the group centered at $\log
  (t_{\mathrm{DF}}/\mathrm{yr}) \approx 8.5$). {\it Top right:} The times for evolution from
  $R_{\mathrm{inf}}$ to coalescence  ($t_{\mathrm{bound}})$. In this regime the orbital evolution is due
  to the combination of the LC scattering, VD in a circumbinary disk, and GW
  emission. {\it Bottom left:} Total evolution times of those systems that evolve from $\sim 1$~kpc to coalescence ($t_{\mathrm{tot}} < t_{\mathrm{Hubble} }$). {\it Bottom right:}
 Typical timescales corresponding to the three mechanisms that operate within $R_{\mathrm{inf}}$. The histograms in this panel are for illustration purposes -- the actual calculation of the MBHBs evolution 
  considered all three processes simultaneously (see Section~\ref{sub:evo})
  } 
\label{fig:dist_tbound}
\end{figure*}
%

\begin{figure*}[t!]
  \centering
    \includegraphics[width=0.49\textwidth]{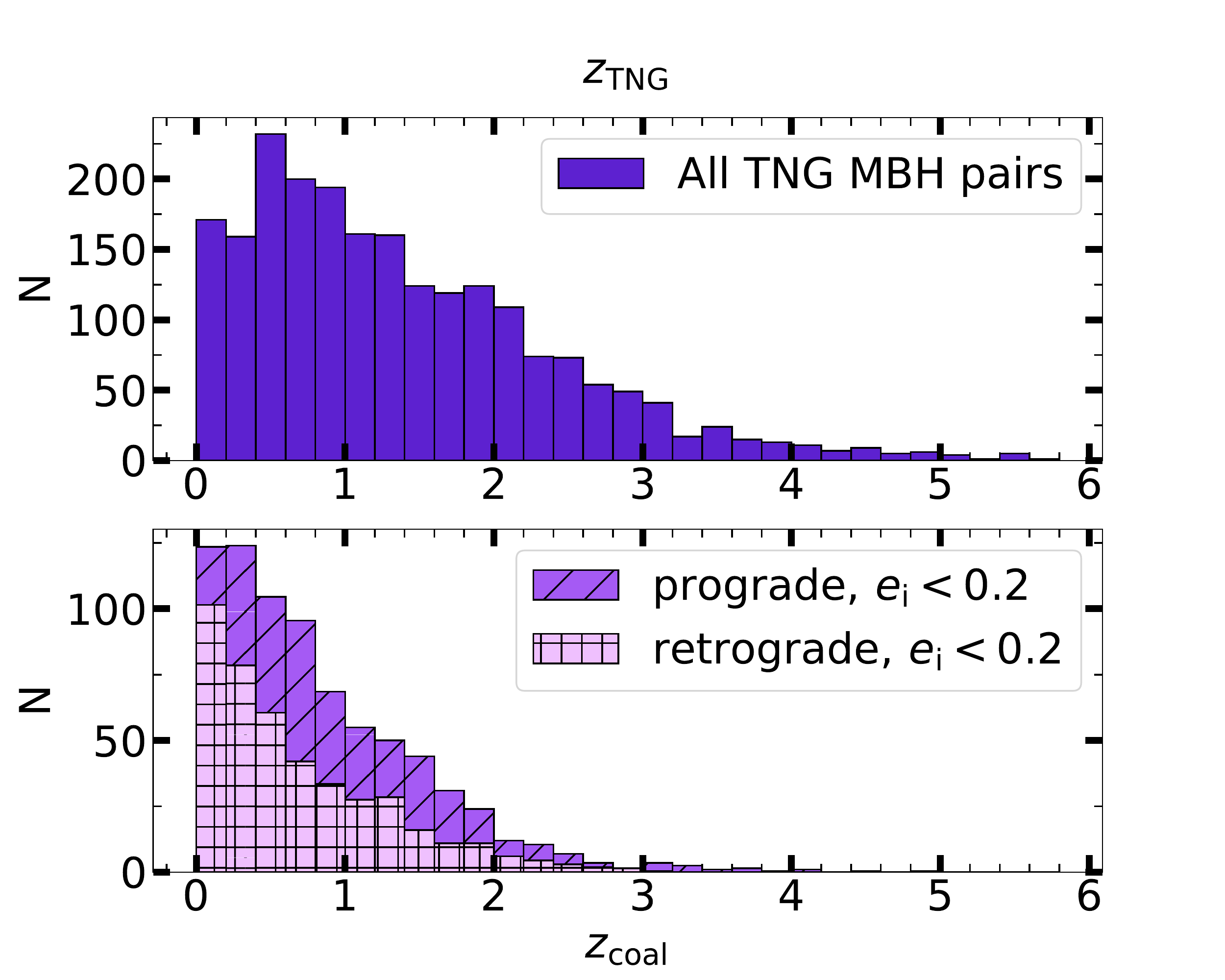}
    \includegraphics[width=0.49\textwidth]{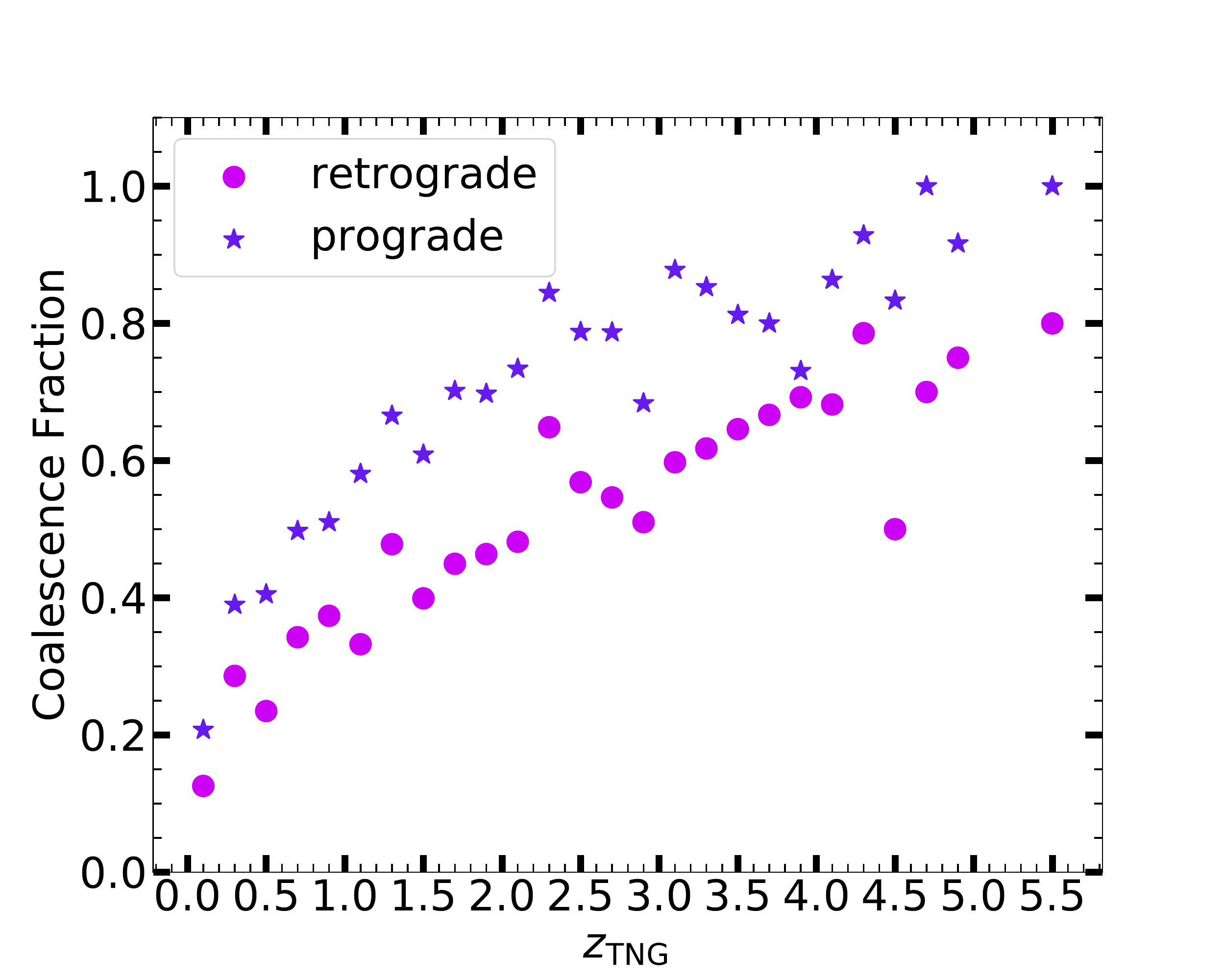}
\caption{\textit{Left:} The top panel shows the redshift distribution
  of 2042 MBH pairs identified in the TNG50-3 simulation
  (Sect.~\ref{sub:TNG}) at the time when they reach the $\sim 1$~kpc
  resolution limit of TNG50-3. This is the starting redshift for all
  evolution calculations in this paper. The bottom panel shows histograms of
  the coalescence redshifts for two sets of orbital configurations:
  $e_i < 0.2$ and prograde (``$/$" histogram) and $e_i < 0.2$ and
  retrograde (``$+$" histogram). In the prograde case $75\%$ of the
    2042 pairs merge by $z=0$, while only $41\%$ merge in the
    retrograde scenario. In both cases, most mergers occur at
    $z_{\mathrm{coal}} \la 0.4$. \textit{Right:} The fraction of
  systems that coalesce before $z=0$ as a function of $z_{\mathrm{TNG}}$. Systems where the
  \Ms\ is on a prograde orbit (stars) are more likely to merge than
  those on retrograde orbits (circles). Given the long \tevol\ of most
  systems (Fig.~\ref{fig:dist_tbound}), only MBH pairs
  that reach a separation of $\approx 1$~kpc at $z \ga 1$ are likely
  to coalesce before $z=0$.}
\label{fig:dist_z}
\end{figure*}


\subsubsection{Gravitational Waves}
\label{subsub:GW}
The last stage of orbital decay is dominated by GW emission, described following \citet{Peters1964} 
\begin{eqnarray}
  \label{eq:GW1}
\left(\frac{df_{\rm orb}}{dt}\right )_{\mathrm{GW}}=\frac{96\, (2\pi)^{\rm 8/3}}{5c^{\rm 5}}(GM_{\rm chirp})^{\rm 5/3}f_{\rm orb}^{\rm 11/3}\,{\cal F}(e),
\end{eqnarray}
and
\begin{eqnarray}
  \label{eq:GW2}
\left(\frac{de}{dt}\right )_{\mathrm{GW}}=\frac{(2\pi)^{\rm 8/3}}{15c^{\rm 5}}(GM_{\rm chirp})^{\rm 5/3}f_{\rm orb}^{\rm 8/3}\,{\cal G}(e),
\end{eqnarray}
where $M_{\rm chirp}=(m_{\rm 1}m_{\rm 2})^{\rm 3/5} (m_{\rm 1}+m_{\rm 2})^{\rm -1/5}$
is the source frame chirp mass. The factors $\mathcal{F}$ and
$\mathcal{G}$ are
\begin{eqnarray}
  \label{eq:fe}
{\cal F}(e)=\frac{1+73/24e^{\rm 2}+37/96e^{\rm 4}}{(1-e^{\rm 2})^{\rm
    7/2}}
\end{eqnarray}
and
\begin{eqnarray}
  \label{eq:ge}
{\cal G}(e)=\frac{304e+121e^{\rm 3}}{(1-e^{\rm 2})^{\rm 5/2}}.
\end{eqnarray}
%


\subsection{Summary}
\label{sub:summary}
In summary, for each of the 8168 orbital configurations (4 for each of
the TNG50-3 derived galaxy models), the decay of the orbit of the \Ms\ is
computed from DF forces (Sect.~\ref{subsub:DF}) from a starting
separation of $\approx 1$~kpc to the MBHB influence radius
$R_{\mathrm{inf}}$. Within that radius, the evolution of the \Ms\ is
  calculated from the combination of LC scattering
  (Sect.~\ref{subsub:LC}), VD forces in a circumbinary disk
  (Sect.~\ref{subsub:VD}), and GW emission
  (Sect.~\ref{subsub:GW}). The sum of the latter three  processes determines
  the orbital decay below $R_{\mathrm{inf}}$ as
  $(df_{\mathrm{orb}}/dt)_{\mathrm{total}}\approx(df_{\mathrm{orb}}/dt)_{\mathrm{LC}}+(df_{\mathrm{orb}}/dt)_{\mathrm{VD}}+(df_{\mathrm{orb}}/dt)_{\mathrm{GW}}$\footnote{This approximation is valid because usually one process dominates the evolution of the MBH pair and because the time-step of the calculation ($5\%$ of the orbital period) is significantly shorter than the inspiral timescale (Fig.~1).}. The evolution of the eccentricity in each time step is assumed to be due
to the dominant mechanism only (given by the largest $df_{\mathrm{orb}}/dt$ in that particular time-step). The
  calculation ends, and the MBHs are deemed to have coalesced, when the orbital separation reaches
  $6GM_{\mathrm{bin}}/c^2$.  We do not consider the evolution of the MBHB orbit due to the influence of any additional
MBHs that may be transported to the galactic nucleus after a subsequent galaxy merger \citep{Hoffman2007, Amaro2010, Ryu2018, Bonetti2019}. The impact of this
assumption is discussed in Section~\ref{sect:discuss}. 



\section{Timescales for Evolution of MBH Pairs from kpc Scales to Coalescence}
\label{sec:time}
Figure~\ref{fig:dist_tbound} shows the distribution of evolution times
found from the suite of 8168 MBH orbits. The histogram in the upper-left panel plots the range of
times, \tevol, found in the DF stage, when the separation drops from $\sim
1$~kpc to $R_{\mathrm{inf}}$. The majority of the studied
configurations have timescales of $t_{\mathrm{DF}} \ga 1$~Gyrs, consistent
with the times found by LBB20a for motions dominated by gaseous
DF.

Indeed, we find that $35$\% of the models never reach coalescence,
with the vast majority of these stalling in the DF phase (the hatched
region on the right-hand side of the \tevol\ histograms). These are
models in which the gaseous DF force is minimized because of the nature of
orbit, leading to very lengthy decay times. This can be seen by
examining the stalled fractions for the individual orbital
configurations: $15$\% (prograde, $e_i < 0.2$), $37$\% (prograde, $0.8
< e_i < 0.9$), $46$\% (retrograde, $e_i < 0.2$) and $41$\%
(retrograde, $0.8 < e_i < 0.9$). The configurations with the largest
stalled fractions are the ones where the \Ms\ will likely have a large
relative velocity relative to the gas disk, diminishing the
effectiveness of DF (LBB20a).

\begin{figure*}[t]
  \centering
    \includegraphics[width=0.49\textwidth]{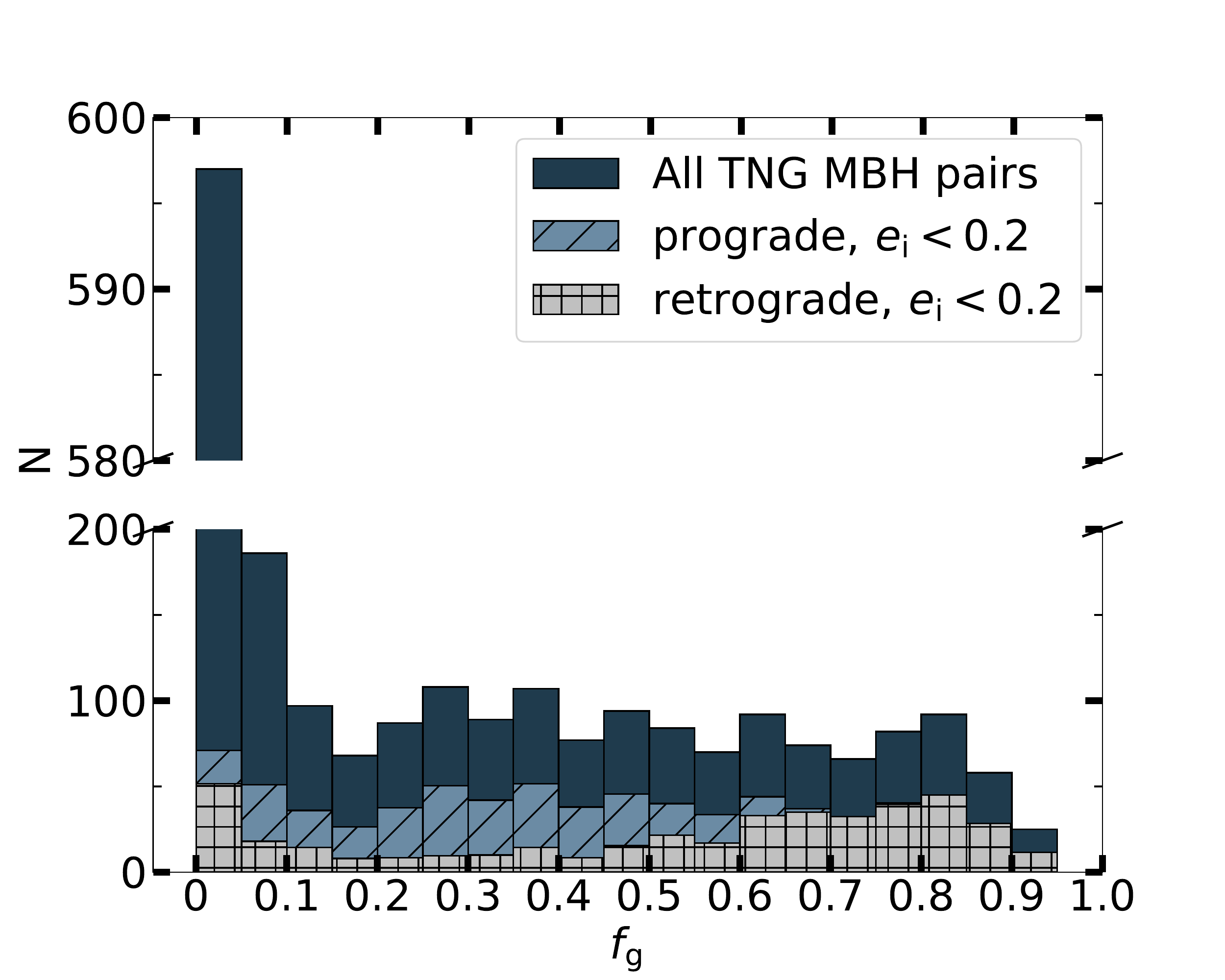}
    \includegraphics[width=0.49\textwidth]{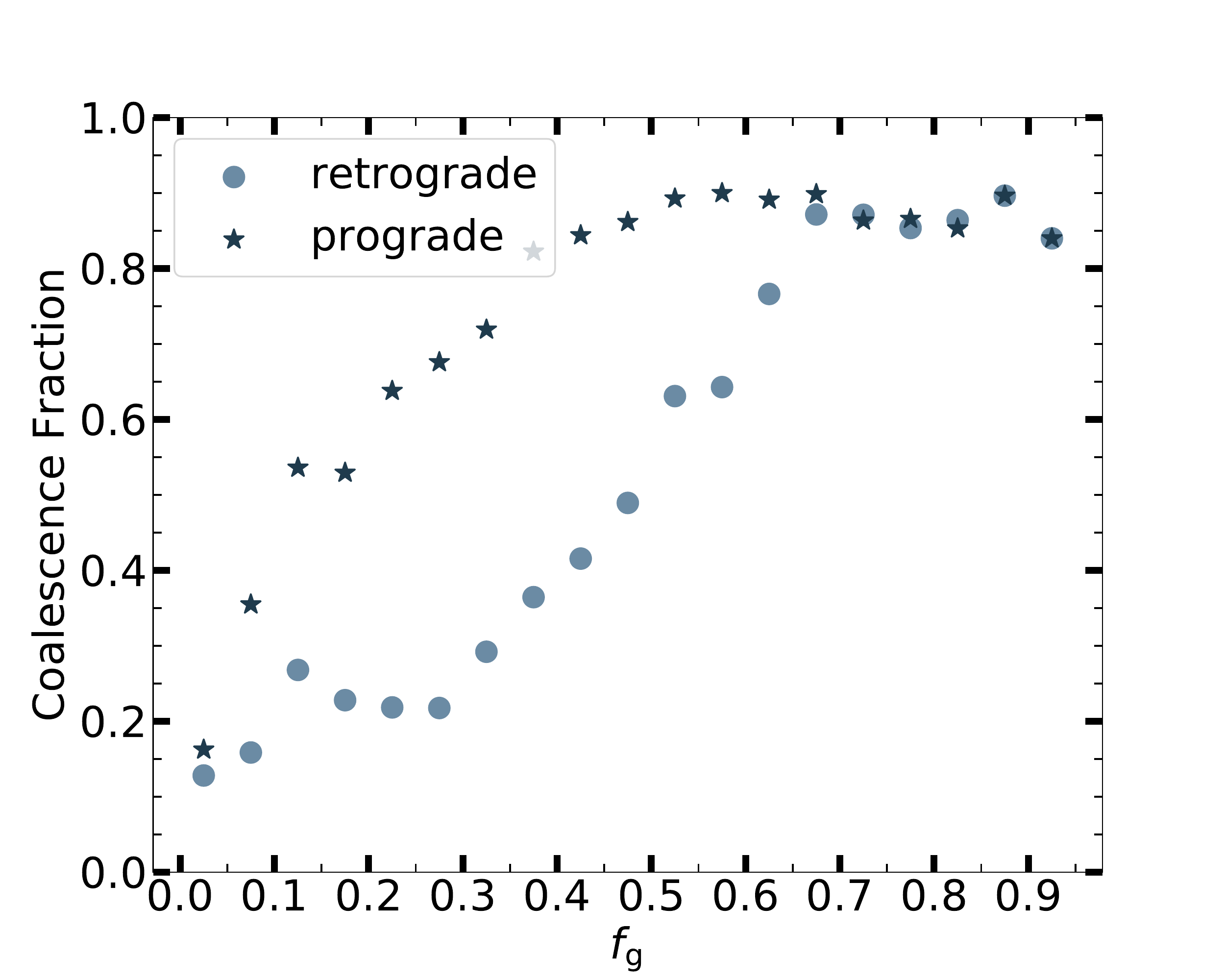}
\caption{\textit{Left:} The solid histogram shows the parent
  distribution of \fg\ for the 2042 post-merger galaxy models characterized
  from the TNG50-3 simulation (Sect.~\ref{sub:TNG}). The other two
  histograms show the \fg\ distributions of galaxies with
  MBHBs that coalesce before $z=0$. The \Ms\ in shown models have
  $e_i < 0.2$, and are on either prograde (``$/$") or
  retrograde orbits (``$+$"). \textit{Right}: The coalescence fraction
  as a function of \fg\ computed from all 8168 MBH orbital
  configurations, in prograde (stars) or retrograde
  (circles) motion. MBH mergers occur more frequently in gas
  rich galaxies ($f_g > 0.2$) because of the stronger gaseous DF forces (LBB20a).}
\label{fig:dist_fg}
\end{figure*}

The remaining 25 MBHBs that do not merge in a Hubble time, as shown by the hatched bars around $\log\,(t_{\mathrm{DF}}/\mathrm{yr}) \approx 8.5$ in Figure~\ref{fig:dist_tbound}, skip the DF-dominated stage and directly enter the LC dominated stage when the calculation starts. Even though they are relatively gas rich (with gas fractions $\gtrsim 0.6$), their spatially extended gas and stellar distributions have densities too low to extract sufficient orbital energy from the sMBHs within $t_{\mathrm{Hubble} }$.

The distribution of timescales from $R_{\mathrm{inf}}$ to
coalescence, denoted as $t_{\mathrm{bound}}$, is shown in the upper-right
histogram. These timescales span four orders of magnitude, from $\sim10^4$ to $10^8$ years, and are thus much shorter than \tevol. As a result, the distribution of the total
evolution time, from about 1\,kpc to coalescence ($t_{\mathrm{tot}}$,
shown in the lower-left) looks nearly identical to the
\tevol\ distribution. It follows that for a vast majority of systems the DF phase is the
most important mechanism in determining the rate of the MBHB coalescences. Stating this differently: the measurement of the rate of MBHB coalescences should provide direct constraints on the rate of evolution of the MBH pairs in merger galaxies due to DF, a point which we discuss in Section~\ref{sub:implications}.

Although the orbital evolution within $R_{\mathrm{inf}}$ is actually computed
using the combination of three mechanisms (LC, VD and GW), it is
interesting to calculate how long would any one of them dominate if it operated by itself (i.e., by neglecting the two sub-dominant mechanisms). These distributions are shown in the lower-right panel of
Fig.~\ref{fig:dist_tbound}, and we emphasize that this way of
separating the different evolution times is only for illustration
purposes. The histograms show that time spent in the LC dominated stage is on average longer than that spent in the VD or GW dominated stage. 

There are 45 model galaxies ($2\%$ of the TNG50-3 derived systems)  with
stellar but no gas components. The evolution times of the sMBHs in
these models are longer
than $t_{\mathrm{Hubble}}$ since their stellar components do not extend far enough to encompass the sMBH 
orbit. Without DF from gas disks to transport them, these sMBHs stall at radii  outside the stellar bulge. 
Moreover, there are another 128 galaxies ($6\%$ of
the TNG50-3 derived systems) with very extended gas and stellar distributions and high orbital eccentricities. The
initial semi-major axes of the MBHBs in these galaxies are already smaller than
$R_{\mathrm{inf}}$,  meaning that these sMBHs skip the DF-dominated stage and
directly enter either the LC- or VD-dominated stages when the calculation
starts. Most of these MBHBs coalesce before $z=0$, and 25 of them stall at the LC dominated stage, as explained in the third paragraph of this section.

Figure~\ref{fig:dist_z} shows how the range in $t_{\mathrm{tot}}$ maps
onto a redshift distribution of MBH mergers. The redshift distribution
shown in the top left panel are in terms of $z_{\mathrm{TNG}}$, the
redshift of the MBH pair when it reaches the TNG50-3 resolution limit of $\approx
1$~kpc. This is the parent distribution at the time when the evolution
calculations started. This distribution peaks at $z_{\mathrm{TNG}}
\sim 0.5$. The hatched histograms in the bottom left panel represent the
coalescence redshift distribution for MBHBs in prograde/retrograde
orbits with $e_i<0.2$ that coalesce before $z=0$. The peak of these
distributions moves towards $z=0$ indicating that most of the MBH
mergers will be at $z \la 0.4$. However, as mentioned above, a large
fraction of the simulated MBH pairs will not merge within a Hubble
time. The right panel of Fig.~\ref{fig:dist_z} shows that the fraction of
systems that coalesce before $z=0$ increases significantly
as a function of the initial redshift, $z_{\mathrm{TNG}}$. In fact, if a prograde MBH
pair at $z_{\mathrm{TNG}} \approx 1$ has a separation larger than $\approx 1$~kpc,
then it has $\la 50$\% chance of reaching coalescence before
$z=0$. This redshift threshold increases to $z_{\mathrm{TNG}} \approx
3$ for pairs on retrograde orbits, because of their much longer
\tevol\ (LBB20a).

\section{The Effect of Galactic Properties on the MBHB Coalescence Rate}
\label{sec:merger_rate}

Here, we describe how properties of the post-merger galaxy affects the
likelihood of MBHB coalescence (Sect.~\ref{sub:galactic}). The integrated
MBHB coalescence rate predicted for the four different orbital
configurations is presented in Sect.~\ref{sub:rates}. 

\begin{figure*}[t]
  \centering
    \includegraphics[width=0.49\textwidth]{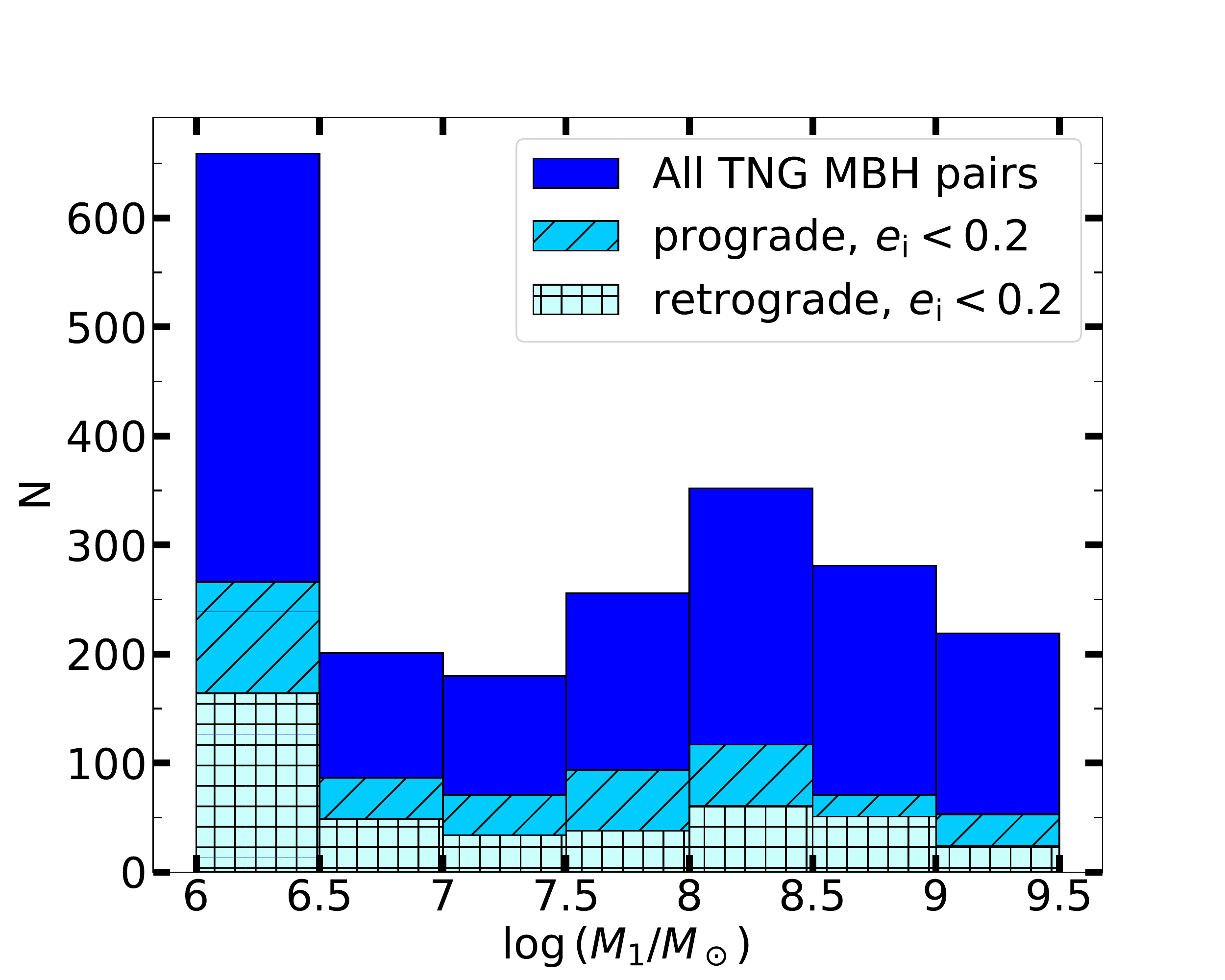}
    \includegraphics[width=0.49\textwidth]{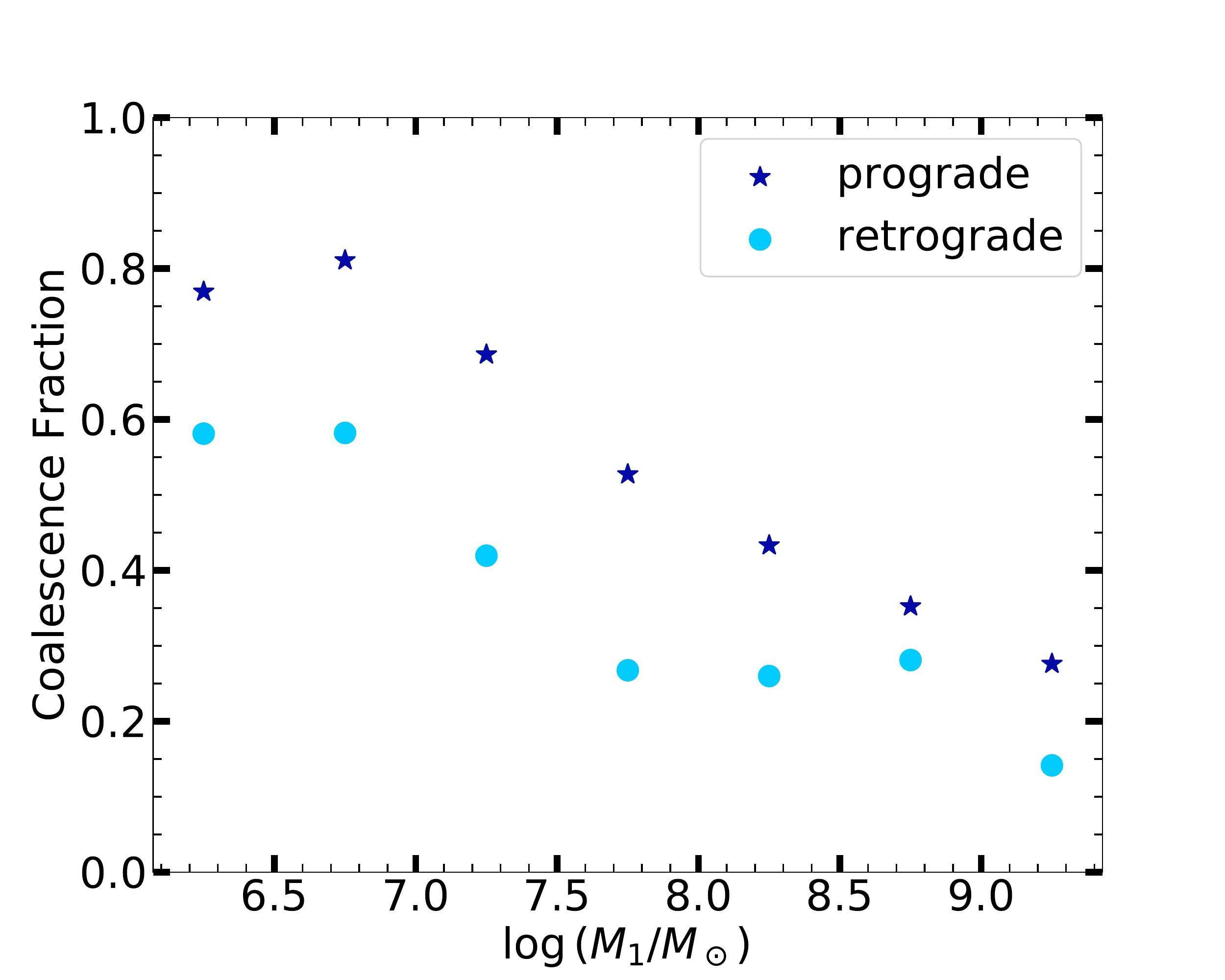}
\caption{As in Fig.~\ref{fig:dist_fg}, but now showing the effect of
  the mass of the \Mp. MBH mergers are more likely for smaller \Mp\ because systems with low \Mtot\ on average have relatively high \fg\ in the TNG50 simulations. The high density gas disk ensures that gaseous DF efficiently decays the orbit of the sMBH.}
\label{fig:dist_logm1}
\end{figure*}

\subsection{MBHB coalescence fraction as a function of $f_g$, $M_1$, $M_{\rm sb}$ and $q$}
\label{sub:galactic}
As discussed in the previous section, whether a MBH pair
eventually coalesces following a merger is determined by the time spent in the DF stage
between $\sim 1$~kpc and $R_{\mathrm{inf}}$. Therefore, the chances that
a coalescence will occur depends most strongly on the properties of the
host galaxy that impact the DF evolution timescale at these
distances. These properties are the gas fraction (\fg), the
mass of the \Mp\ ($M_1$), the bulge mass (\Msb), and the MBH mass ratio
(\q)\footnote{See LBB20a for the effect of these properties on the DF force.}.\\

\noindent (a) {\it Gas fraction, $f_g$ --} The left panel of
Figure~\ref{fig:dist_fg} shows the parent population of \fg\ in the
galaxy models constructed from the TNG50-3 simulation
(Sect.~\ref{sub:TNG}). Overplotted are two
\fg\ distributions of post-merger galaxies that host a \Ms\ in a
low $e_i$ orbit and in which the MBHB merges before $z=0$. 
These distributions show that the MBHB mergers are more likely to
happen when $f_g > 0.2$, especially when the \Ms\ is on a prograde
orbit. This is a consequence of the increased efficiency of gaseous DF
forces in decaying the orbit of the \Ms\ (LBB20a). In particular,
retrograde orbits rarely result in a high coalescence fraction
because of the high relative velocity between the \Ms\ and the gas disk
lowers the effectiveness of DF (LBB20a).

This conclusion is reinforced by examining the coalescence fraction
from all 8168 orbital evolution models as seen in the right panel of
Fig.~\ref{fig:dist_fg}. As \fg\ increases from zero, the coalescence
fraction for prograde orbits (stars) increases and reaches a maximum at $0.5 \leq f_{\rm g} \leq
0.6$, after which it
becomes roughly flat for larger \fg. At larger $f_{\rm g}$, the difference in coalescence fraction caused
by the orbital direction nearly vanishes due to the \Ms\ moving
subsonically in the high density and high temperature gas.\\



%
%

\noindent (b) {\it Mass of the primary MBH, $M_1$ --}
Figure~\ref{fig:dist_logm1} shows the dependence of the coalescence
fraction on the mass of the
\Mp. The peak of the mass distribution for parent \Mp s  from TNG50-3 is at $10^{\rm 6} \leq
M_{\rm 1} \leq 10^{\rm 6.5} M_{\rm \odot}$ and the coalescence fraction is
highest for galaxies with these pMBHs. This is a result of the fact
that systems with low \Mtot\ often have medium to high \fg\ in the
TNG50-3 simulations. For example, MBH pairs with $M_{\mathrm{bin}} \sim 10^7 M_{\rm \odot}$ ($ \sim 10^8 M_{\rm \odot}$) tend to have $f_g \sim 0.4$ ($\sim 0.2$), and higher values of \fg\ lead to more efficient evolution
via gaseous DF (Fig.~\ref{fig:dist_fg}).\\

\begin{figure*}[t]
  \centering
    \includegraphics[width=0.49\textwidth]{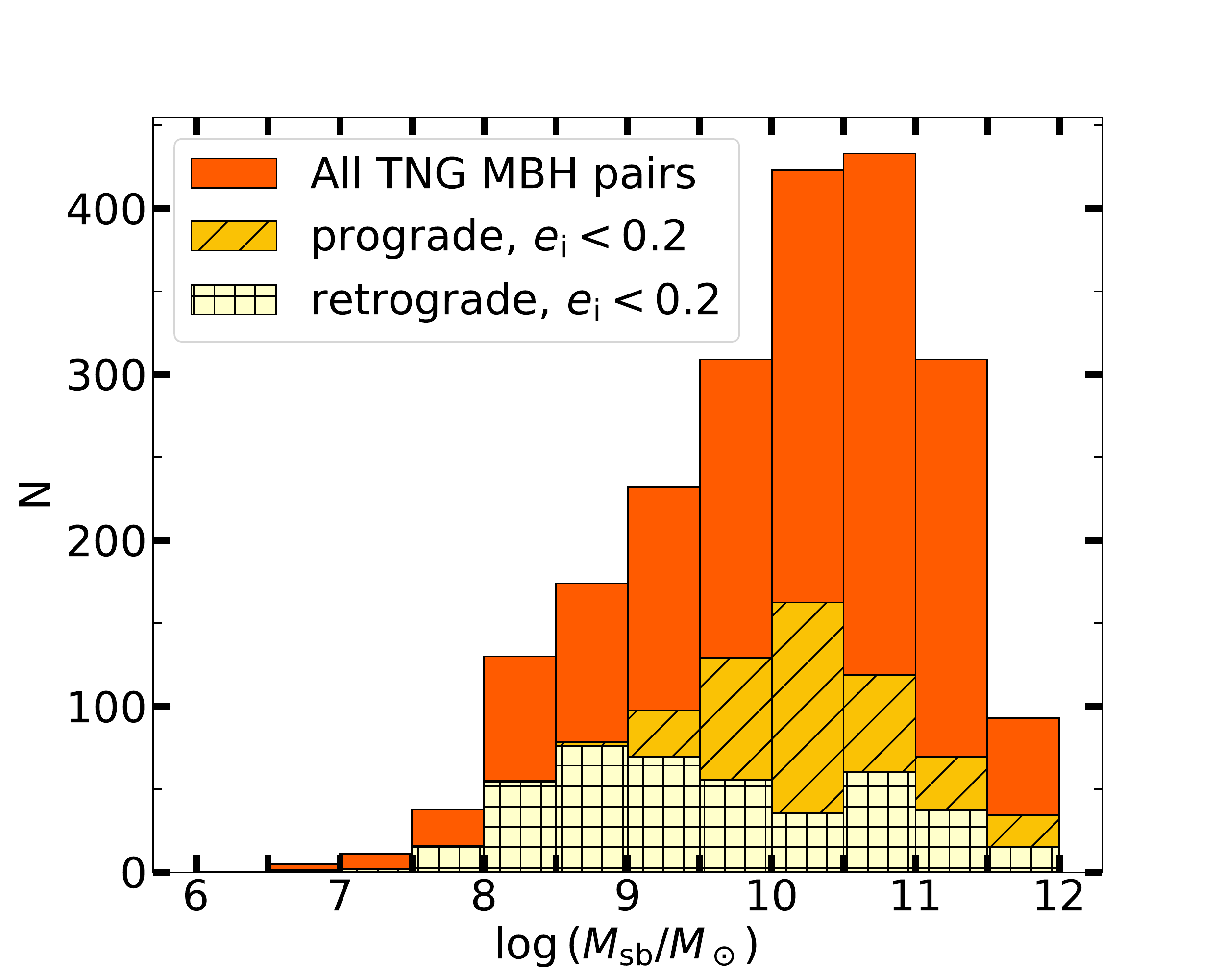}
    \includegraphics[width=0.49\textwidth]{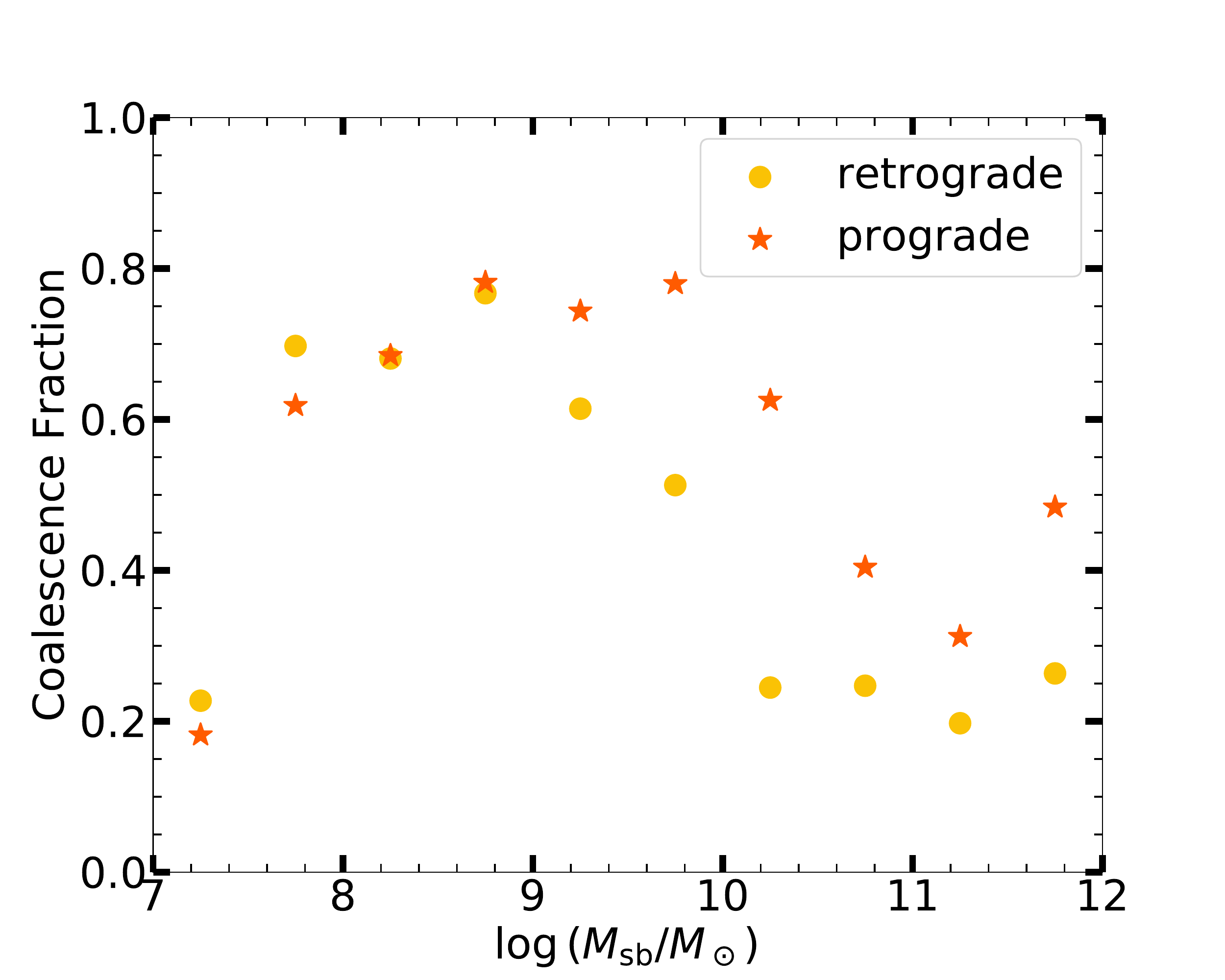}
\caption{As in Fig.~\ref{fig:dist_fg}, but now showing the effect of
  the mass of the stellar bulge, \Msb. The coalescence fractions are
  largest for bulges with $9 \la \log(M_{\mathrm{sb}}/M_{\odot}) \la
  10$, as these masses correspond to values of \fg\ where
  gaseous DF is efficient. }
\label{fig:dist_logmb}
\end{figure*}

\noindent (c) {\it Bulge mass, $M_{\rm sb}$ --} Figure~\ref{fig:dist_logmb} shows the dependence of the coalescence
fraction on the mass of the stellar bulge, \Msb. The parent distribution peaks at $10
\leq \log(M_{\rm sb}/M_{\odot}) \leq 11$, but the coalescence
fractions are largest for $9 \leq \log (M_{\rm sb}/M_{\rm \odot}) \leq 10
$. These merger fractions are largely influenced by the relationship
between \fg\ and \Msb\ (Sect.~\ref{sub:galaxymodel}).  When the bulge
mass is smaller than $10^{\rm 9} M_{\rm \odot}$, the galaxy models
tend to have high \fg\ (and thus, high gas densities and sound speeds) on average,   
so their \Ms s move subsonically, and experience reduced gaseous DF.
For \Msb\ between $10^{\rm 9}$ and $10^{\rm 10} M_{\rm \odot}$, \fg\ becomes smaller and gaseous DF efficiently
decays the orbit of the \Ms\ (see also Fig.~\ref{fig:dist_fg}). When
$M_{\mathrm{sb}} \ga 10^{\rm 10} M_{\rm \odot}$, \fg\ falls to low enough
values that gaseous DF again becomes inefficient.\\

\begin{figure*}[t]
  \centering
    \includegraphics[width=0.49\textwidth]{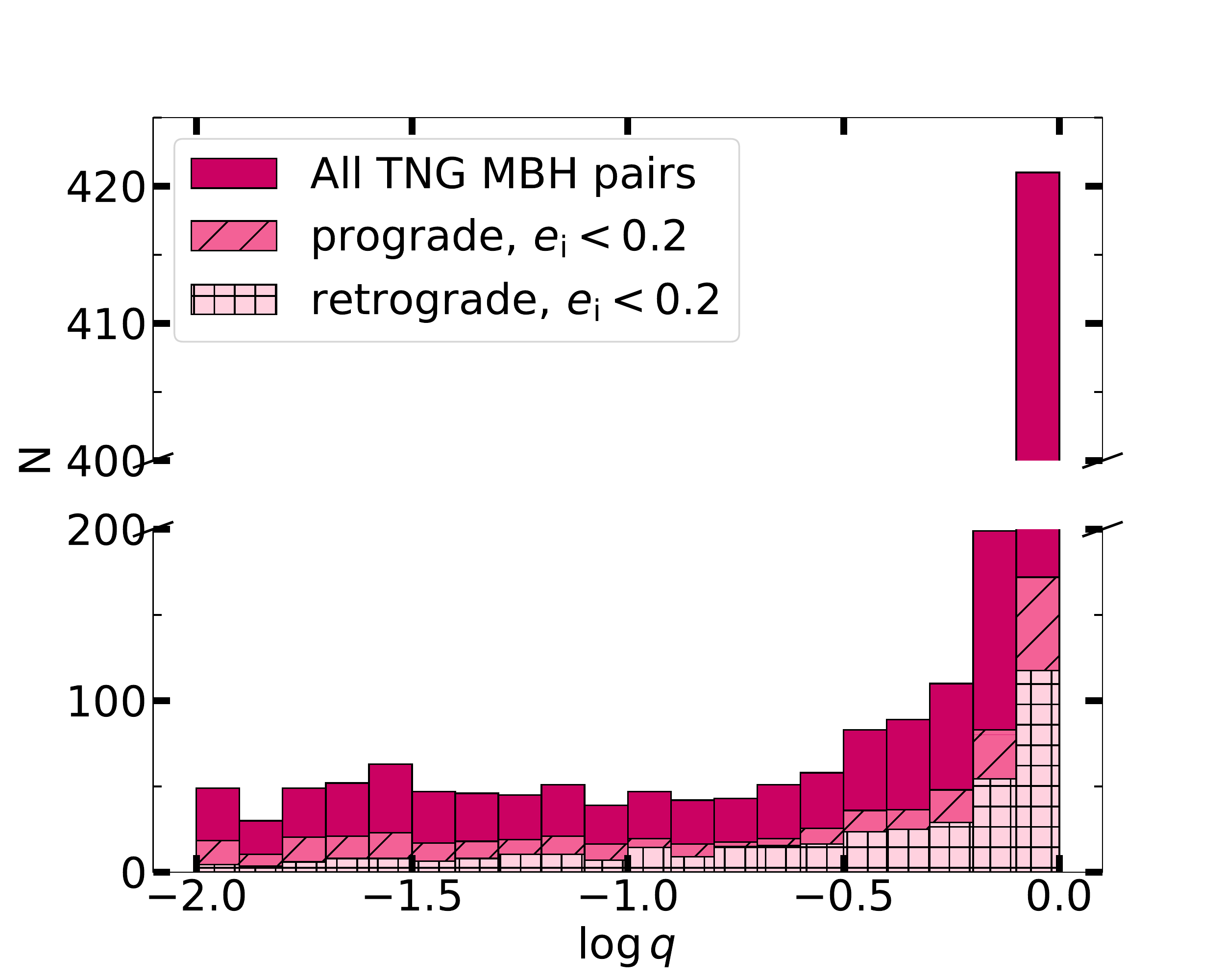}
    \includegraphics[width=0.49\textwidth]{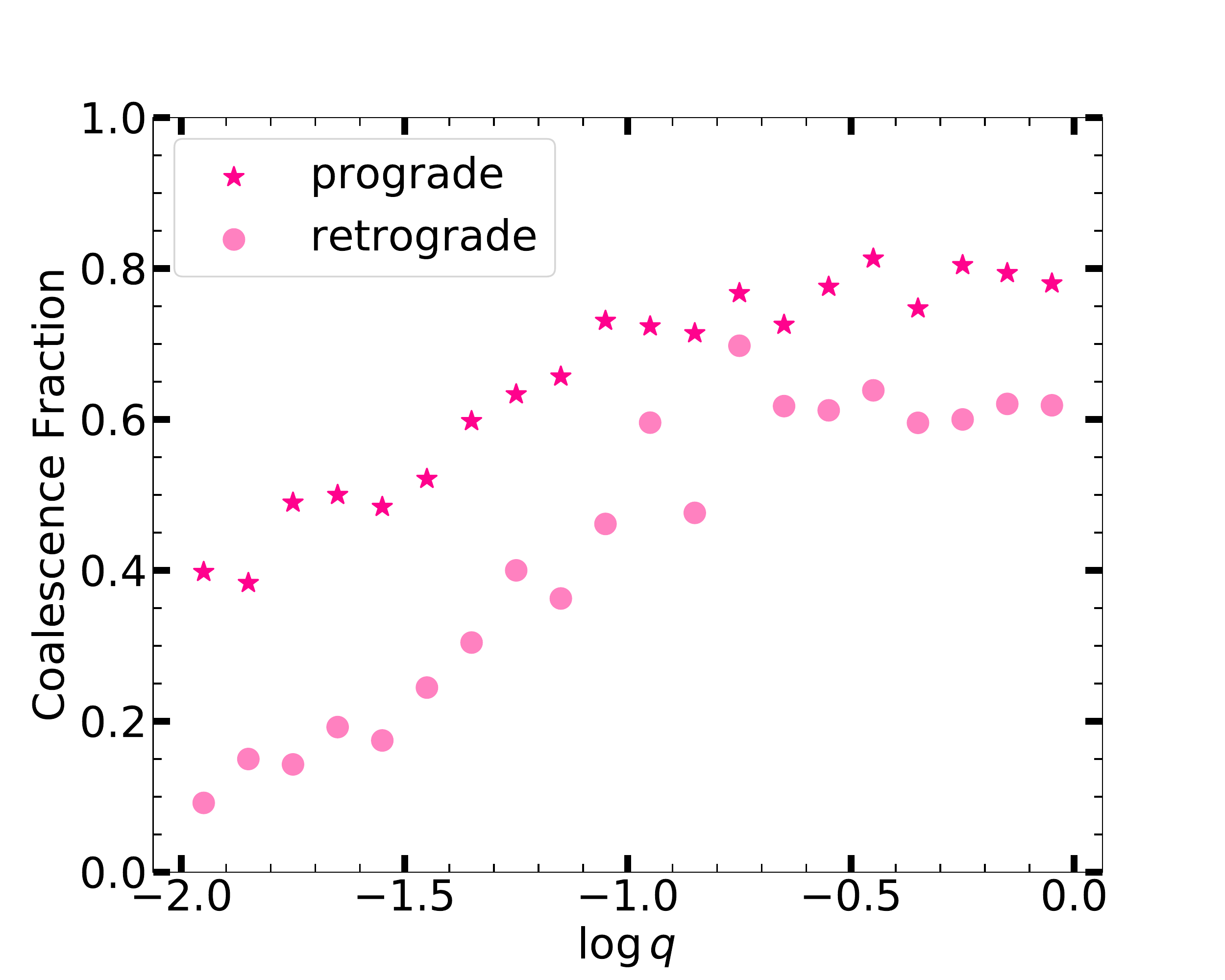}
\caption{As in Fig.~\ref{fig:dist_fg}, but now showing the effect of
  the MBH pair mass ratio, \q. The largest coalescence
  fractions occur when $q \ga 0.1$ because higher mass \Ms s
  experience stronger DF forces (LBB20a). }
\label{fig:dist_q}
\end{figure*}

\noindent (d) {\it MBHB mass ratio, $q$ --} Figure~\ref{fig:dist_q} shows the dependence of the coalescence
fraction on the mass ratio of the MBH pair, \q. The parent distribution of MBH pairs taken
from the TNG-50 simulation peaks at $q \sim 1$, but the coalescence
fraction is roughly constant (at 70-80\%) for $q \ga 0.1$. These
relatively high merger fractions arise because the DF force is larger
(and inspiral time shorter) for higher mass sMBHs (LBB20a; their Figure 11). Interestingly, there is a
significant fraction of mergers ($\sim 50\%$) even at a relatively low
mass ratio of $\log q \approx -1.5$ when the \Ms\ is in a prograde
orbit. This indicates that secondary MBHs on prograde orbits will have
a decent probability of merging before $z=0$ over a very broad range
of $q$.


%
%
 
\subsection{The MBHB coalescence rate}
\label{sub:rates}

The results in the previous section can now be used to compute the
integrated MBHB coalescence rate from $z=0$ to $7.86$ (the largest `merger' redshift in TNG50-3) for the four
different orbital configurations of the \Ms. The integrated
coalescence rate ($dN_{\mathrm{mer}}/dt$) is defined as the total
number of MBHBs that reach coalescence before $z=0$ per unit observer
time \citep{Haeh1994},
\begin{equation}
\label{eq:merger_rate}
\frac{dN_{\rm mer}}{dt}=\int_{0}^{\infty}n_{\rm mer}(z)\times \frac{4\pi c d^{\rm 2}_{L}(z)}{(1+z)^{\rm 2}}{\rm d}z,
\end{equation}
where $d_L$ is the luminosity distance to redshift $z$ and $n_{\rm
  mer}(z)=dn/dVdz$ is the comoving number density of merging
MBHBs. For each of the four orbital configurations, $n_{\mathrm{mer}}$
is calculated from binning all coalescences in unit intervals of $z_{\rm coal}$ and
dividing it by the total comoving TNG volume, ($51.7$~cMpc)$^{\rm 3}$. The resulting values of $dN_{\mathrm{mer}}/dt$ calculated from
Eq.~\ref{eq:merger_rate} are listed in the second column of
Table.~\ref{tab:rate}.

\begin{deluxetable*}{ccccc}[t!]
\tablenum{1}
\tablecaption{MBHB Coalescence and \textit{LISA} detection rates in
  the absence and presence of radiation feedback (RF)\label{tab:rate}}
\tablewidth{0pt}
\tablehead{
\colhead{Orbital Configuration} & \colhead{Coalescence Rate
  (yr$^{-1}$)} & \colhead{Detection Rate (yr$^{-1}$)} &
  \colhead{Coalescence Rate (yr$^{-1}$)} & \colhead{Detection Rate
      (yr$^{-1}$)}\\ 
 & & & (with RF) & (with RF)
}
\startdata
 prograde and $e_i < 0.2$  & 0.45 &  0.34 & 0.1 &  0.02 \\
 prograde and $0.8 \leq e_i \leq 0.9$ & 0.32 &  0.30 & 0.04 &  0.02 \\
 retrograde and $e_i < 0.2$ & 0.2 &  0.14 & 0.09 &  0.02 \\
retrograde and $0.8 \leq e_i \leq 0.9$ & 0.26 &  0.24 & 0.04 &  0.02
\enddata
\tablecomments{The coalescence rates, $dN_{\mathrm{mer}}/dt$, are
  calculated from Eq.~\ref{eq:merger_rate}. The adopted \textit{LISA}
  detection threshold is SNR$ >8.0$. The effects of
  radiation feedback on the MBHB coalescence and \textit{LISA} detection rates are discussed in Section.~\ref{sec:RF_lisa}.}
\end{deluxetable*}

The largest predicted coalescence rate, $dN_{\rm mer}/dt=0.45$~yr$^{-1}$,
occurs if the sMBHs are on prograde orbits with low
initial eccentricity $(e_i<0.2)$. This drops to
$0.32$~yr$^{-1}$ for MBHs on high $e_i$, prograde orbits. MBHs on high
eccentricity orbits spend large fractions of orbital period moving slowly
at large distances, where the gas density is low. Thus, gaseous DF
forces are less effective in decaying these high eccentricity orbits
(LBB20a). MBH pairs where \Ms s are moving on retrograde orbits have the
lowest coalescence rate of $0.2$~yr$^{-1}$ (low $e_i$) or $0.26$~yr$^{-1}$ (high
$e_i$). These low rates are due to the large relative velocity of the
gas disk and the \Ms, which reduces the effectiveness of the gas DF
force (LBB20a). MBHs with such orbital configurations tend to stay at
large separations for a long time, and many of them do not reach
coalescence before $z=0$.


\section{The Effect of Galactic Properties on \textit{LISA} Detection Rates}
\label{sec:property_lisa}
The previous section described how the MBH coalescence fractions and
rates are impacted by the properties of post-merger galaxies and the
orbital configuration of their \Ms s. We now utilize our MBH evolution
calculations to predict the \textit{LISA} detection rates of
inspiralling MBHBs as a function of the same properties. The calculation of the signal-to-noise ratio (SNR) for a GW event detected by \textit{LISA} is presented in Appendix~\ref{app:GW_emission}. In this work we will assume SNR of 8 as the
detection threshold for \textit{LISA}, as by \citet{Bonetti2019}. 

For each of the four orbital configurations considered in our
calculations, we identify the systems with MBHs that
coalesce before $z=0$. We then evaluate the cumulative SNR in the inspiral phase only using
Eqs.~(\ref{eq:LC1}), (\ref{eq:LC2}), (\ref{eq:gas_drag}),
(\ref{eq:GW1}), (\ref{eq:GW2}), (\ref{eq:snr}), and assuming a four
year \textit{LISA} mission lifetime. We neglect the contribution to the SNR from
the merger and ringdown phase, as it is relatively small and does not
change the predicted \textit{LISA} detection rates
\citep{Bonetti2019}. The resulting \textit{LISA} detection
rates for the four configurations are listed in the 3rd column of
Table~\ref{tab:rate}. Unsurprisingly, the detection rates follow the
same pattern as the total MBHB coalescence rates with the largest
rate ($0.34$~yr$^{-1}$) occurring in systems where the
secondary is on a prograde orbit with low $e_i$, and with retrograde configurations
giving the lowest rates. Interestingly, there is a small difference
between the coalescence and \textit{LISA} detection rates for MBHBs
which began with a large $e_i$. Thus, almost all coalescing systems
with high eccentricity yield a \textit{LISA} detection.



\begin{figure*}[t]
            \includegraphics[width=0.49\textwidth , trim={0 6cm  0 6cm},clip]{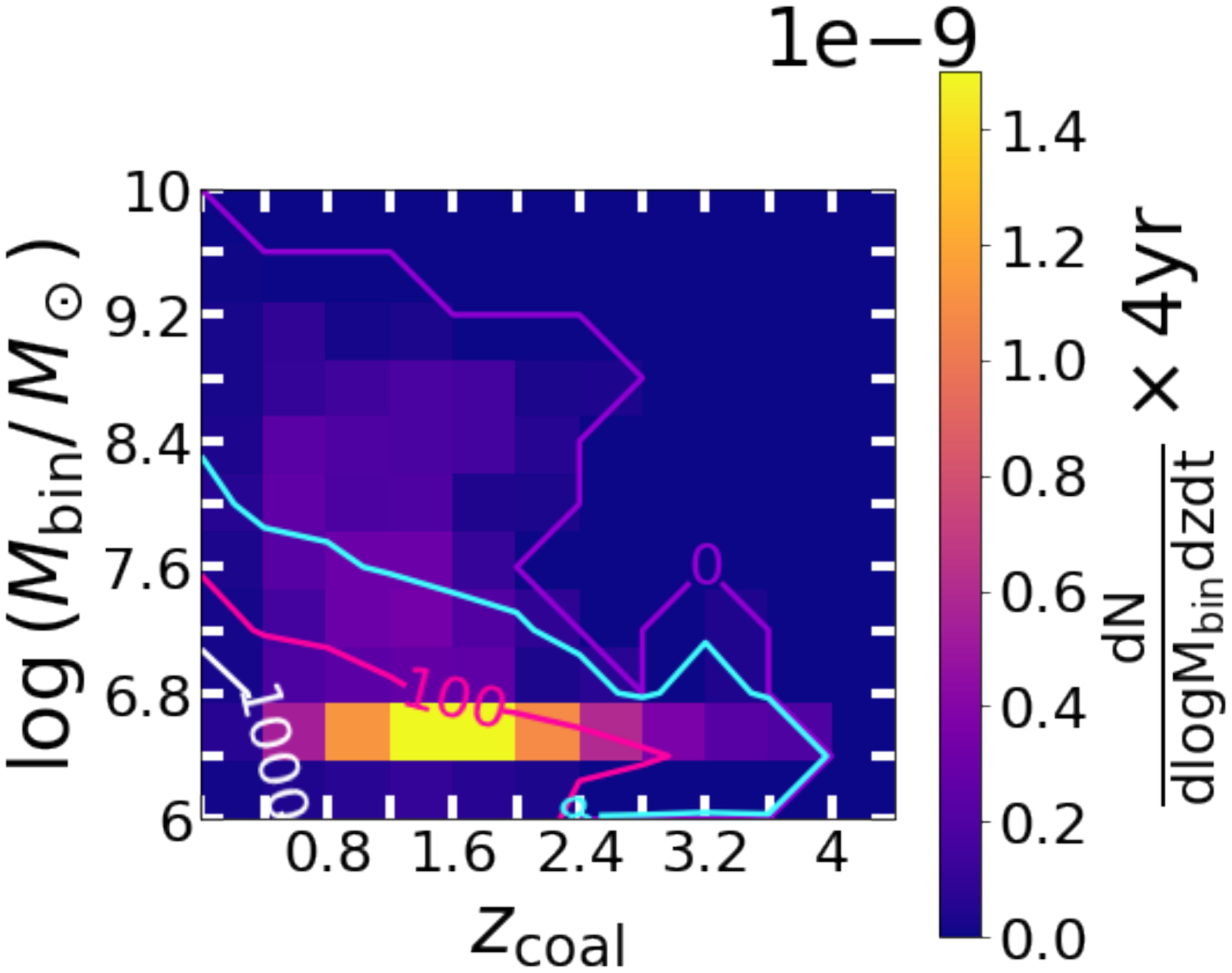}
           \includegraphics[width=0.49\textwidth , trim={0 6cm  0 6cm},clip]{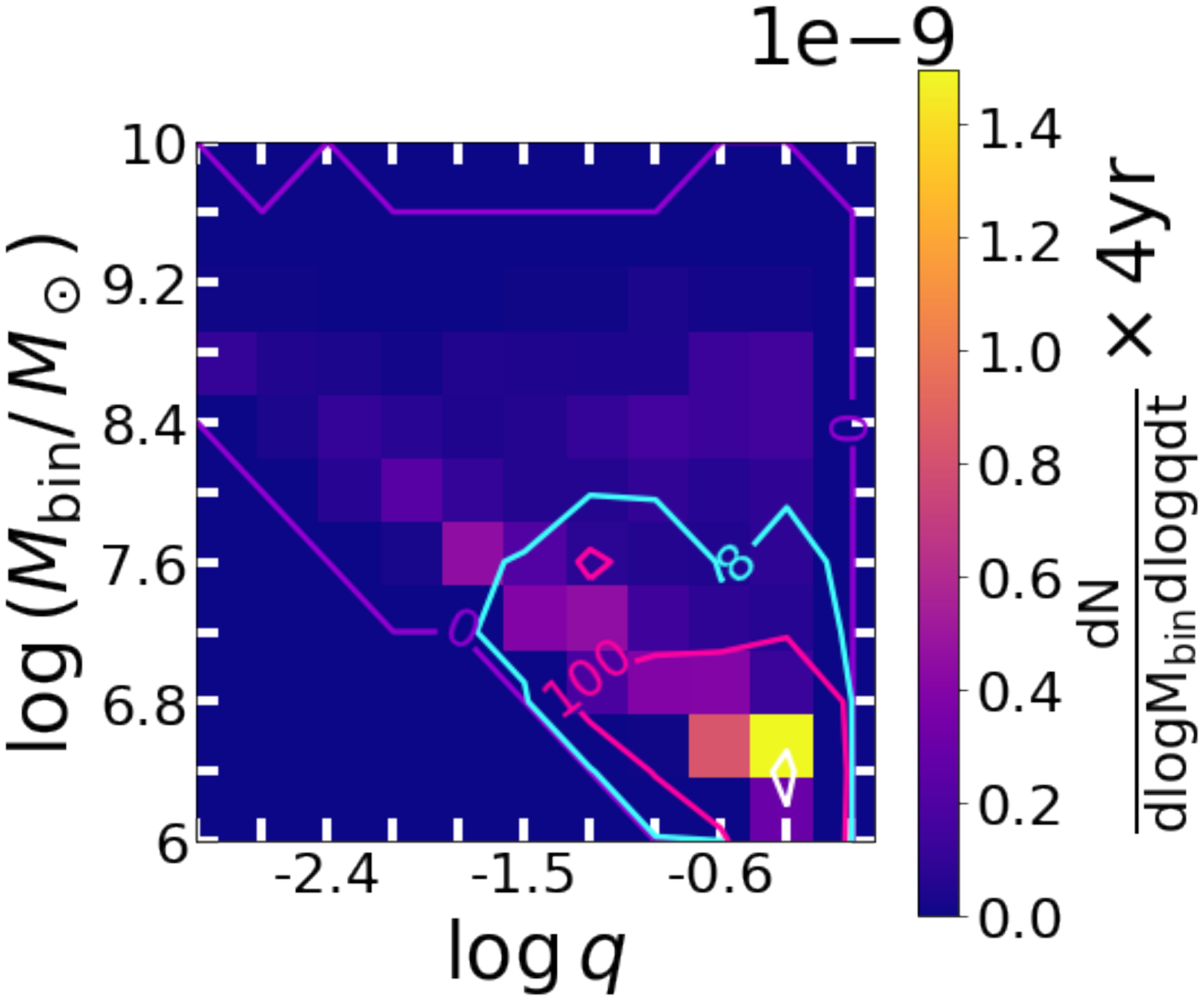}
           \includegraphics[width=0.49\textwidth , trim={0 6cm  0 6cm},clip]{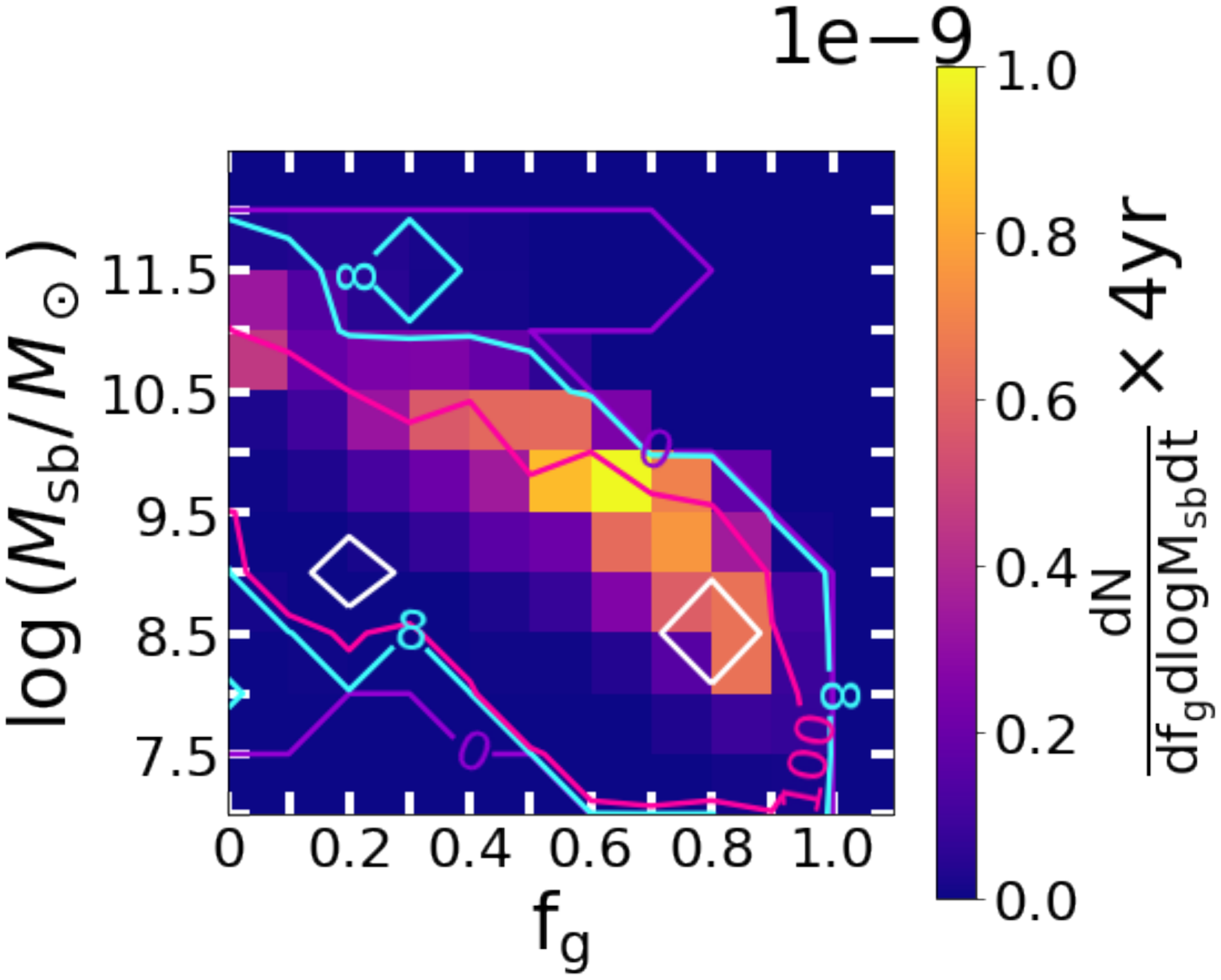}
           \includegraphics[width=0.49\textwidth , trim={0 6cm  0 6cm},clip]{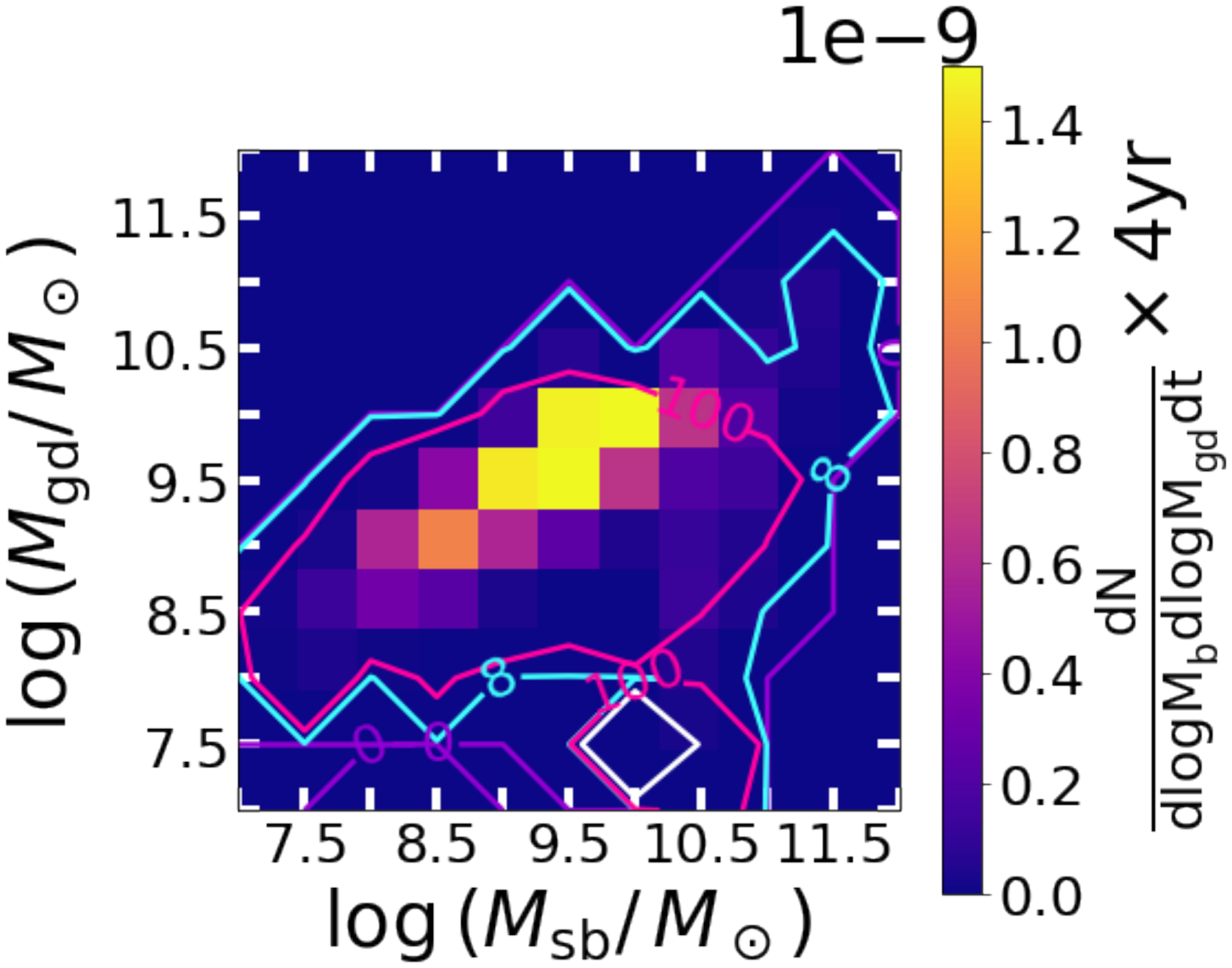}
\caption{Differential number of MBHB mergers over a four-year
  \textit{LISA} mission time
  (shown by the color bar) as a function of the binary mass and
  coalescence redshift (top left), the binary mass and mass ratio (top
  right), the bulge mass and gas fraction (bottom left) and the gas
  disk mass and bulge mass (bottom right). The contours in each panel
  mark the \textit{LISA} SNR: 0 (purple), 8  (cyan), 100 (magenta) and 1000 (white). These panels
 shows the results for the prograde, $e_i < 0.2$ orbital
 configurations. The results are qualitatively similar for the remaining configurations.}
\label{fig:2D_snr}
\end{figure*}

To illustrate how \textit{LISA} detections are
connected to the parameters of the host galaxy models,
Figure~\ref{fig:2D_snr} shows the
two-dimensional differential number of all mergers expected in four years
of \textit{LISA} mission time for four different combinations of the
galaxy properties. Overplotted on each figure are contours of the
\textit{LISA} SNR. The figure makes use of the prograde, low $e_i$ set
of models. The three remaining orbital configurations result in qualitatively similar distributions with different normalizations.

The
upper-left panel shows the differential number of mergers as a
function of $\log (M_{\rm bin}/M_{\odot})$ and $z_{\rm coal}$. The number of
coalescences in four years is highest in systems with $1.2\leq
z_{\rm coal}\leq 2$ and $ 6.4 \leq \log (M_{\rm bin}/M_{\odot}) \leq
6.8$. The detected SNR also peaks on the low MBHB mass end and is
higher at smaller $z_{\rm coal}$. MBHBs with total mass $\sim 10^{\rm 6}
M_{\rm \odot}$ are ``louder" than others because
\textit{LISA} is most sensitive in the frequency range corresponding
to their mass. The SNR of MBHBs at low redshifts is large simply
because of their proximity to the Earth. Taking both the coalescence frequency and
SNR into account, we expect that most \textit{LISA} detections should
originate from $1.2 \leq z \leq 2$, and have SNR $\sim
100$. Detections of sources with SNR $>1000$ and $z<0.4$ would be
extremely rare.

The upper-right panel of Fig.~\ref{fig:2D_snr} shows the
differential number of mergers as a function of $\log M_{\rm bin}$
and $\log q$. The number of mergers in four years is the highest in
systems with $0.5 \leq q \leq 1$ and $ 6\leq \log (M_{\rm bin}/M_{\odot}) \leq
6.8$. The SNR distribution peaks in the same location in the $\log (M_{\rm bin}/M_{\odot})-\log q$ parameter space, hence we expect these MBHBs to
be the loudest and most frequently detected \textit{LISA} sources. 

The differential number of mergers as a function
of $\log (M_{\rm sb}/M_{\odot})$ and \fg\ is shown in the lower-left panel. The
number of mergers in four years is highest in systems with $f_{\rm g} \approx 0.6$ and $9.5 \leq \log (M_{\rm sb}/M_{\odot}) \leq 10
$, as seen earlier in Figs.~\ref{fig:dist_fg}
and~\ref{fig:dist_logmb}. The \textit{LISA} SNR$=8$ contour
envelopes all values in this parameter space. Therefore, the probability of a
\textit{LISA} detection is largely independent of the \fg\ and
\Msb\ of the host galaxy.


Finally, the differential number of mergers as a function of \Mgd\ and
\Msb\ is shown in the lower-right panel of Figure~\ref{fig:2D_snr}. The
coalescence frequency is the highest in systems with $9.5 \leq \log
(M_{\mathrm{gd}}/M_{\odot}) \leq 10.5$ and $9 \leq \log (M_{\rm
  sb}/M_{\odot}) \leq 10.5$, as this combination of gas disk and bulge
mass leads to the highest \fg\ and coalescence rates. Again, the
SNR$=8$ contour envelopes these points, indicating that these mergers
will nearly all be detected by \textit{LISA}.

In summary, \textit{LISA} should detect most of the MBH pairs in the TNG50-3 simulation that form before $z_{\mathrm{coal}} \ga1.2$. This is because the
systems that merge this quickly are ones with efficient gaseous DF,
which largely depends on the gas fraction of the host galaxy. In the
TNG50-3 simulation, higher \fg\ is typically associated with lower mass
MBHs, placing these mergers squarely in the prime sensitivity range of
\textit{LISA}. As a
result, we predict that \textit{LISA} detections of inspiralling MBHBs
will be concentrated in systems with binary masses in the range of
$6.4 \la \log (M_{\mathrm{bin}}/M_{\odot}) \la 6.8$ and mass ratios of
$q\approx 0.5 -1$, residing in
galaxies with $f_g \approx 0.6$, and gas disk and bulge masses of $\sim10^{\rm 9}-10^{\rm 10.5} M_{\rm
  \odot}$. Furthermore, these detections are expected to be
characterized by SNR $\sim 100$.  

\begin{figure*}
\centering
  \includegraphics[width=0.33\textwidth]{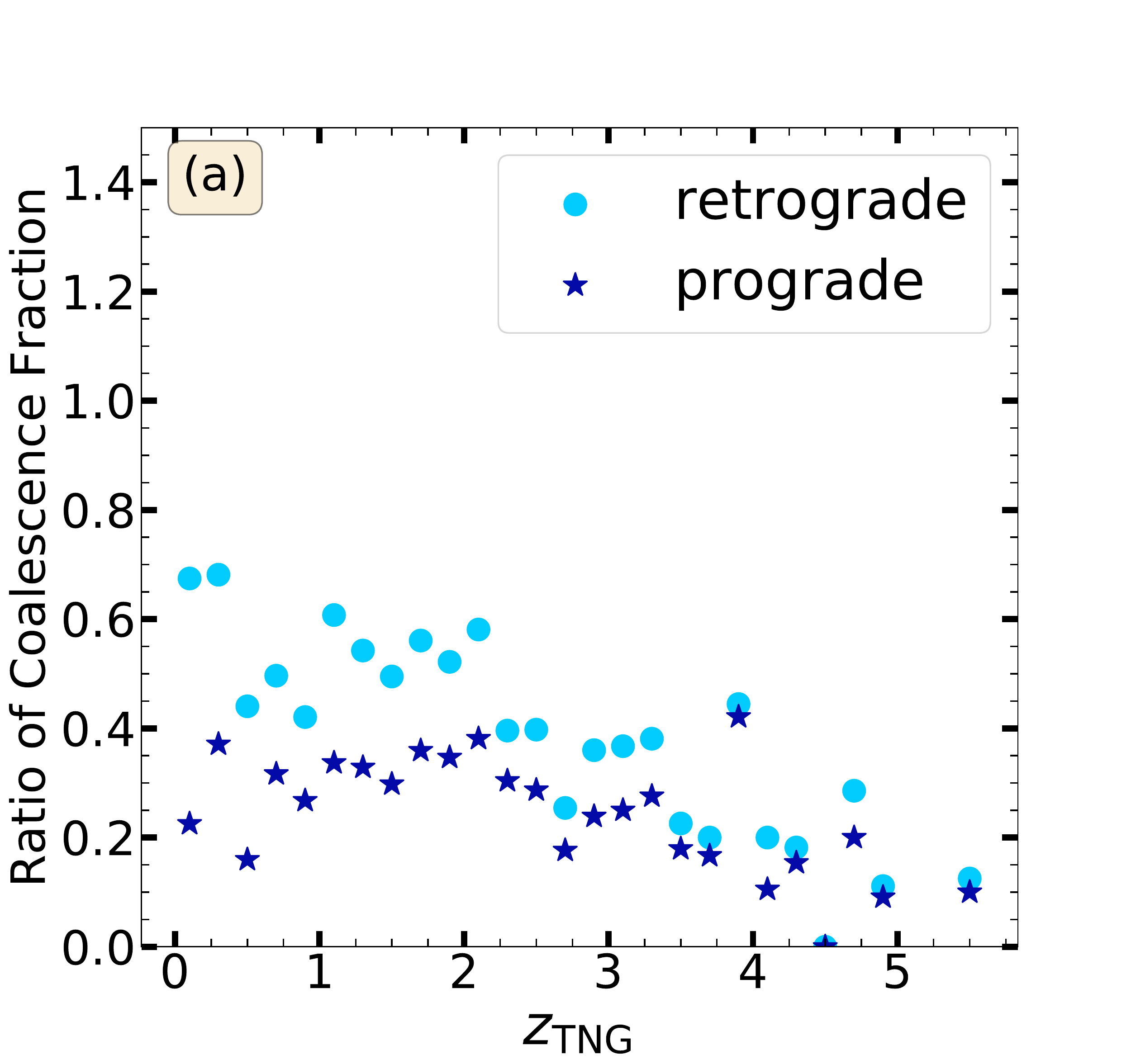}
  \includegraphics[width=0.33\textwidth]{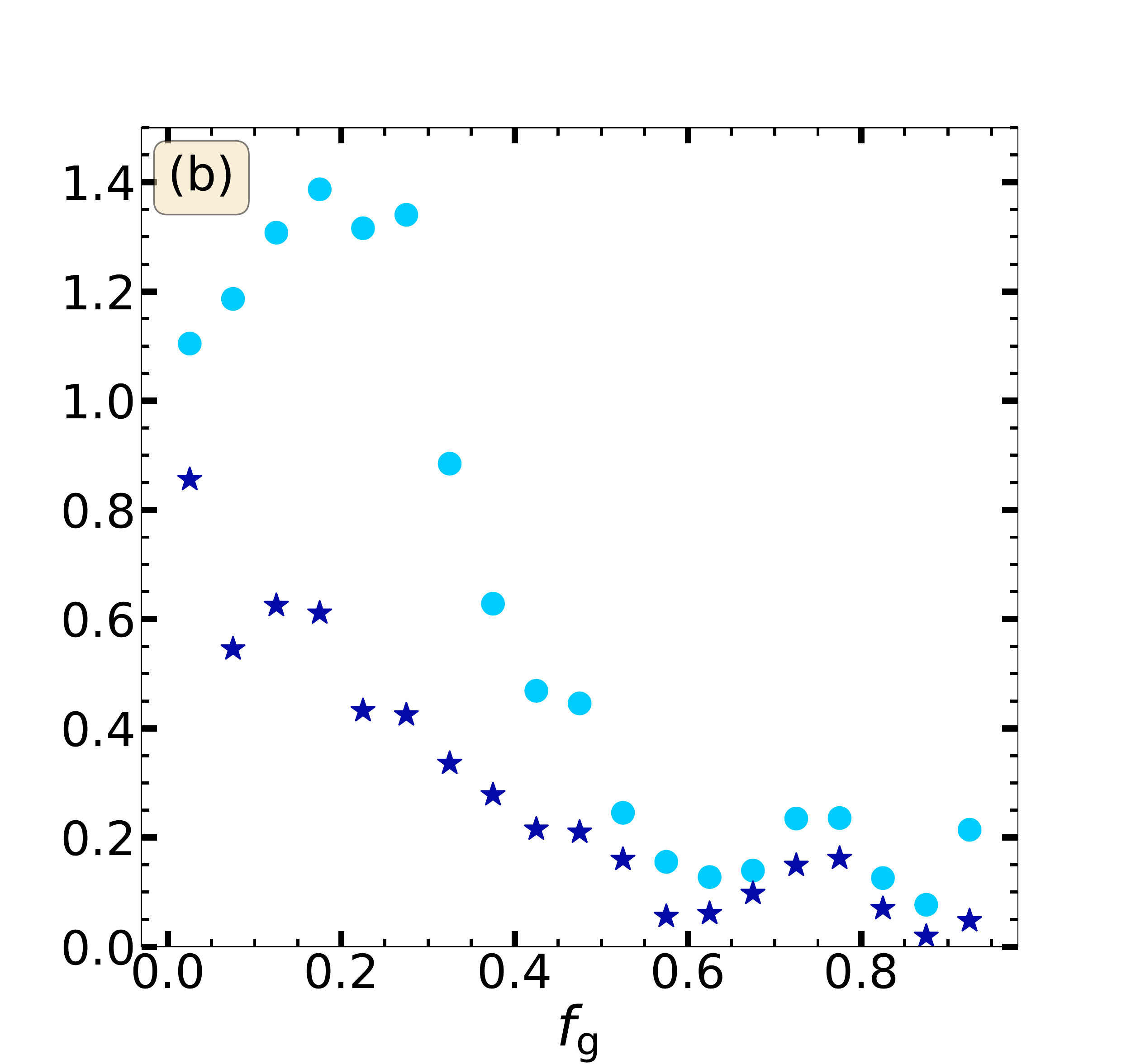}
\includegraphics[width=0.33\textwidth]{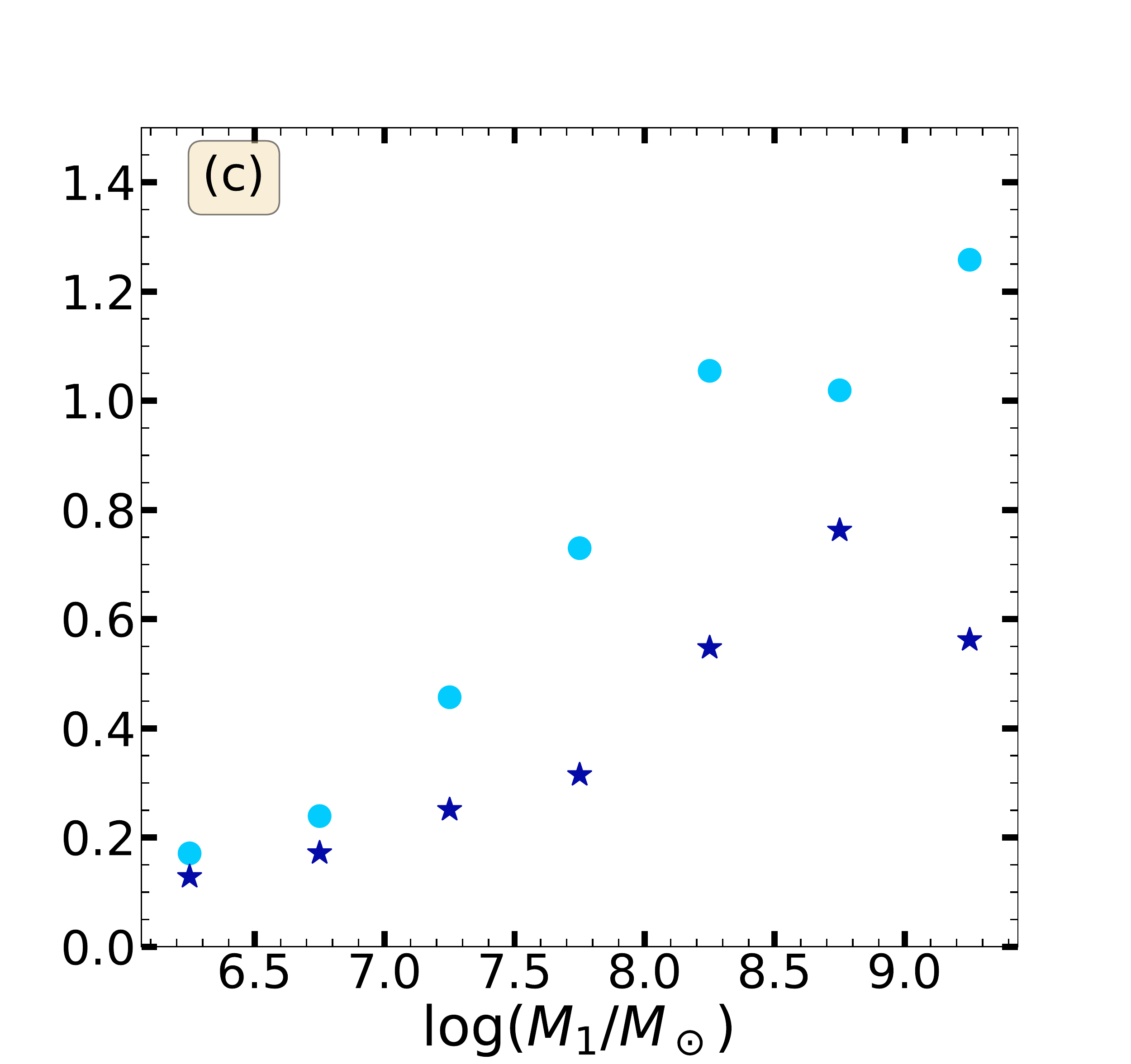}\\
\includegraphics[width=0.33\textwidth]{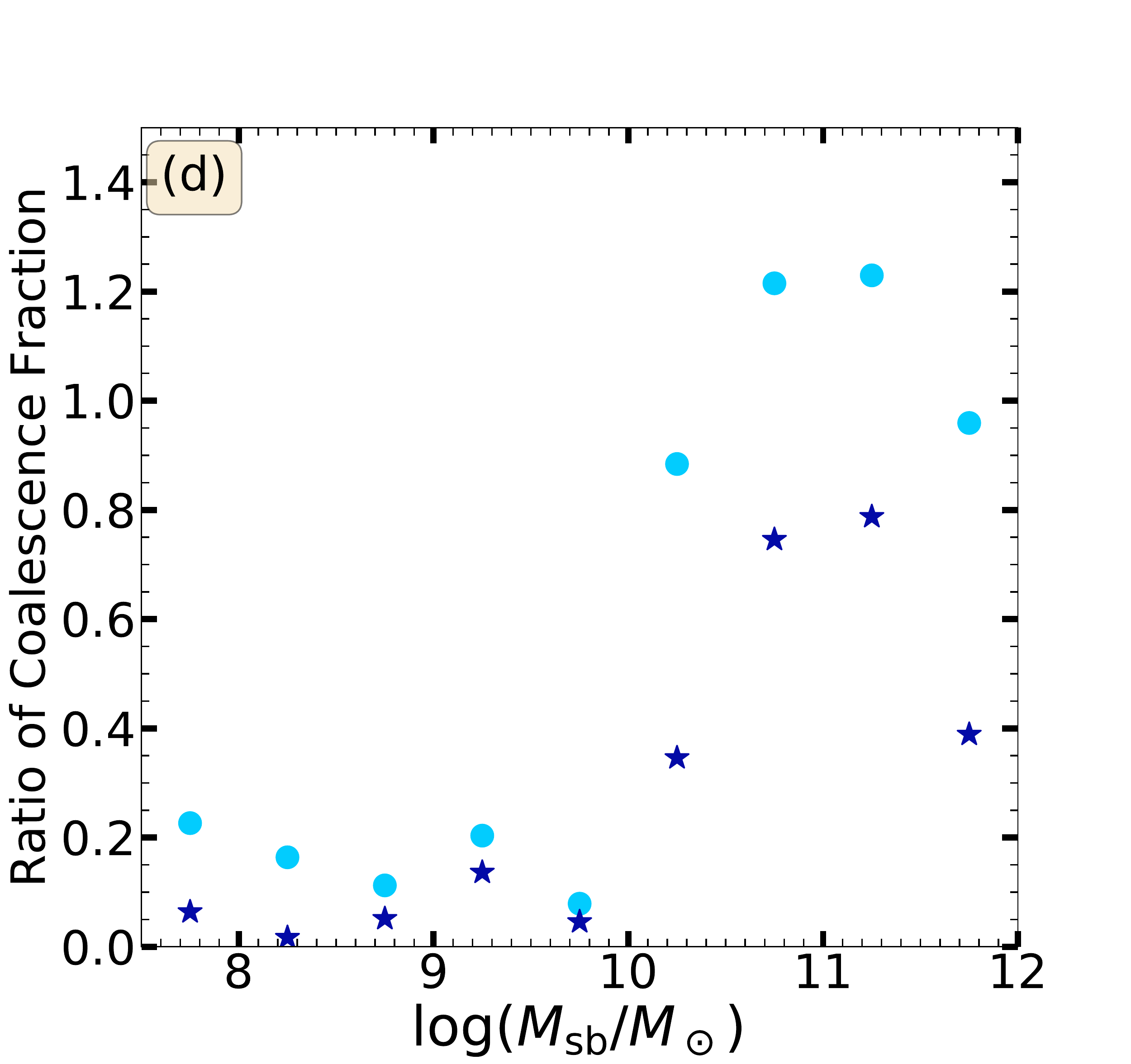}
\includegraphics[width=0.33\textwidth]{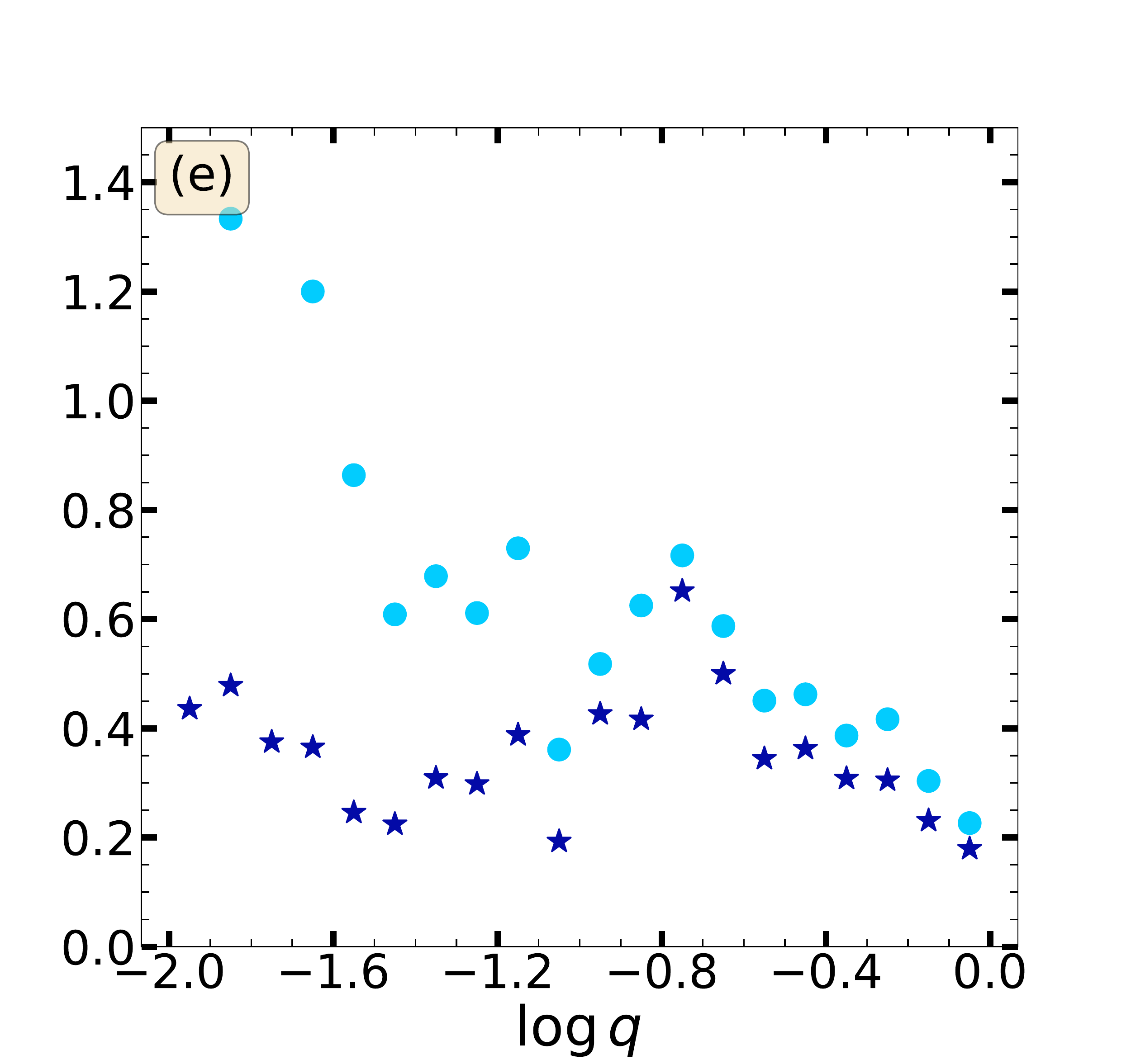}
\caption{The effect of RF on the coalescence fraction of MBHBs, shown as the ratio of the fraction of coalesced MBHBs in the calculation with RF and the fraction of coalesced MBHBs in the calculation without RF. The stars (circles)
  mark ratios for MBHBs in prograde (retrograde) orbits. RF
  significantly reduces the fraction of MBHB coalescences in these
  TNG50-3 derived galaxies when the \Ms\ is on a prograde orbit.} 
\label{fig:ratio_RF}
\end{figure*}
%

\section{The Impact of Radiation Feedback (During the Dynamical Friction Phase)}
\label{sec:RF_lisa}

The results presented above show that the MBH coalescence and
\textit{LISA} detection rates predicted from the TNG50-3 galaxies
sensitively depend on the efficiency of the gaseous DF force on the
\Ms. It is therefore worth evaluating any effects that arise during the DF driven evolution that could potentially modify the predictions laid out in the earlier sections. One important effect not taken into account in the calculation of the effect of DF up to this point is that the \Ms s residing in gas-rich hosts will accrete from their surroundings, as they move
through the inner-kpc, and form a hot, ionized bubble around the MBH. Depending on the velocity of
the \Ms\ relative to its gaseous surroundings, and the properties of the gas, the
gaseous bubble may pull the MBH forward in its orbit, diminishing the
effect of gaseous drag, as expected in the absence of radiative feedback \citep{PB2017,G2020,T2020}. This effect, dubbed ``negative DF", slows down the inspiral of sMBHs, increases the time they spend at large
separations, and therefore potentially reduces the number of MBHB
mergers that occur in a Hubble time (LBB20b). Given the
significance of gaseous DF for the predicted merger rates (manifested as a dependence on the gas fraction $f_g$), 
we recalculated the evolution for all MBH pairs drawn from TNG50-3, now taking into account the
effects of radiation feedback (RF) as described by LBB20b. As seen
below, the negative DF significantly alters the expectations for \textit{LISA} detections.

\subsection{The Effects of Radiation Feedback on the Coalescence Rates}
\label{sub:RF_merger_rate}

Figure~\ref{fig:ratio_RF} shows how the coalescence fractions
presented in Sect.~\ref{sub:galactic} change with the inclusion
of RF. Each panel shows the ratio of the coalescence fractions (defined
as the fraction of coalesced MBHBs in the calculation with RF divided 
by the fraction of coalesced MBHBs in the calculation without RF)
for different properties of the model post-merger galaxies. 
As in Figs.~\ref{fig:dist_z}--\ref{fig:dist_q}, we show results for all four
orbital configurations. The panels in  Figure~\ref{fig:ratio_RF} illustrate a substantial reduction in the
coalescence fraction for MBHs on prograde orbits and in gas-rich galaxies (i.e., high \fg, low $M_1$, or
low \Msb; panels b, c and d). In these cases, the coalescence fractions fall to less
than $10$--$20$\% of the values found without RF. 

This dramatic effect
results from the fact that MBHs on prograde orbits are more likely to
have relative velocities that are comparable to the local sound
speed, a necessary condition for negative DF to operate (LBB20b, their
Eq. 1). In contrast, MBHs on retrograde orbits will more likely have
large relative velocities and are therefore less impacted by negative DF. 
Therefore, we find that MBHs on retrograde orbits are more likely to coalesce before
$z=0$ in the presence of RF effects (panel a), a reversal from the findings in Sect.~\ref{sub:galactic}. 
We emphasize that RF only affects MBHBs during the evolution stages dominated by 
gaseous DF  and is not accounted for in the evolution within $R_{\mathrm{inf}}$. Hence, the drop in
coalescence rate of MBHBs is a direct consequence of the increase in \tevol\ caused by
the negative DF. The potential impact of RF on the orbital evolution
at even smaller separations is discussed in Sect.~\ref{sect:discuss}.

\begin{figure*}[t]
            \includegraphics[width=0.49\textwidth , trim={0 6cm  0 6cm},clip]{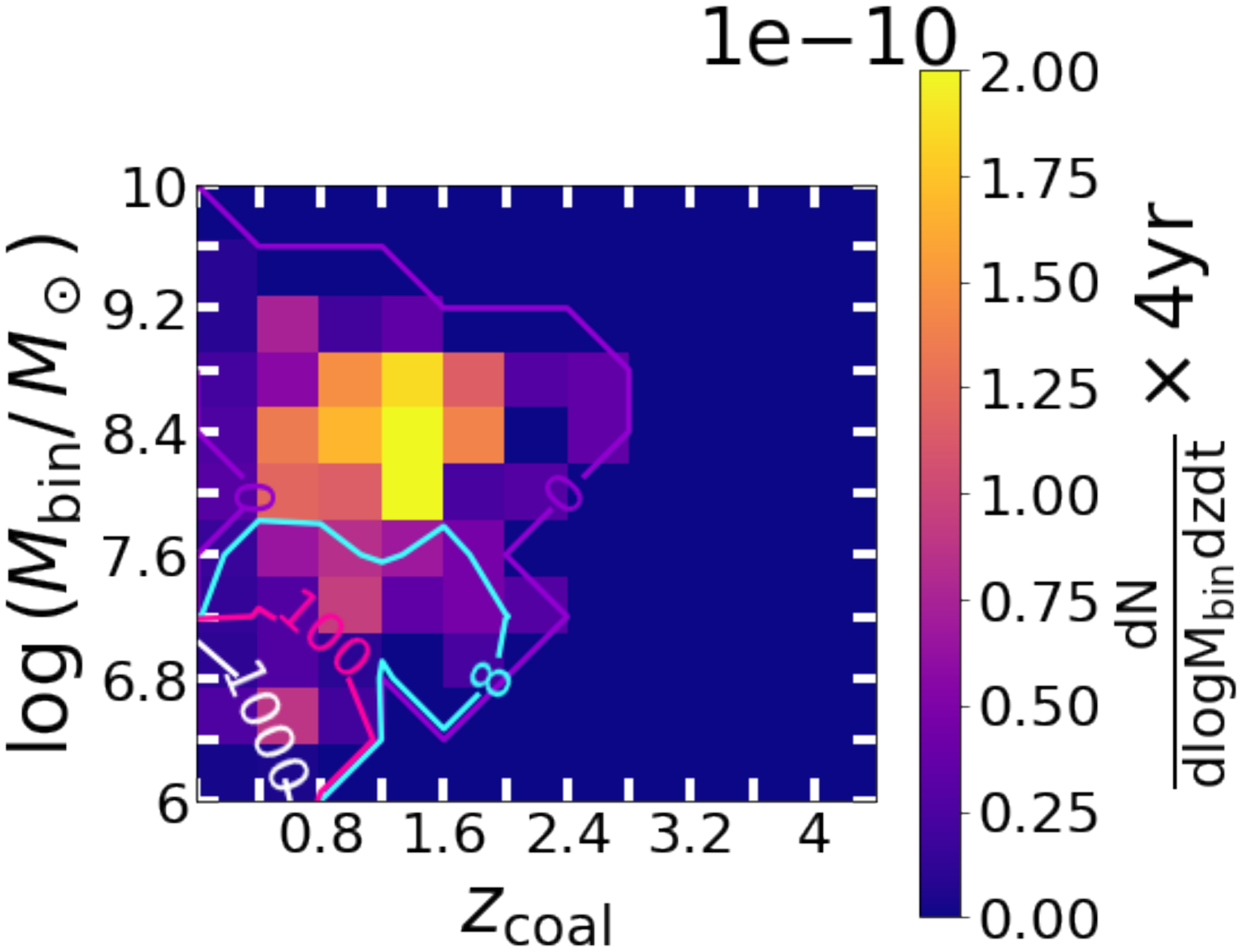}
            \includegraphics[width=0.49\textwidth , trim={0 6cm  0 6cm},clip]{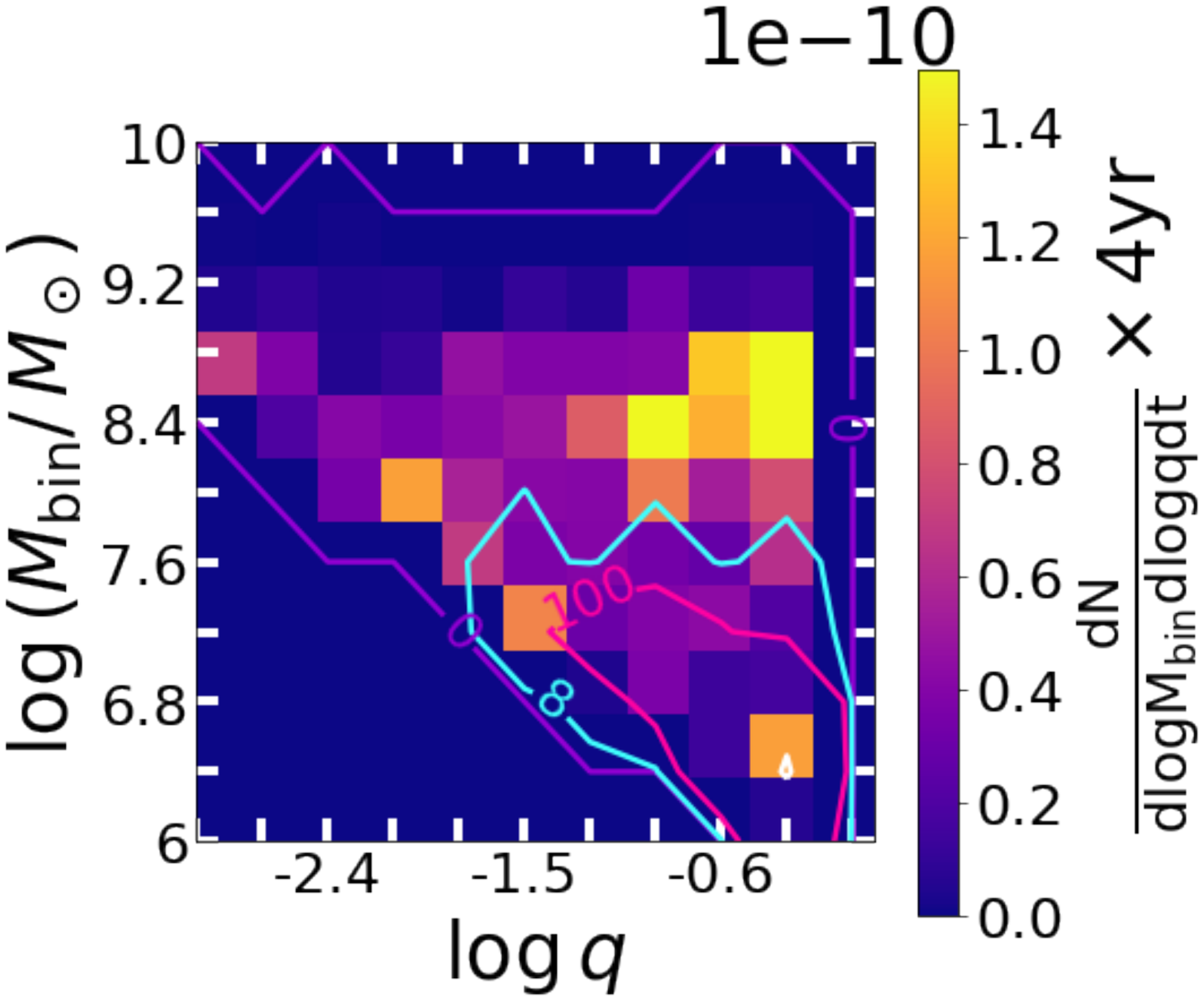}
            \includegraphics[width=0.49\textwidth , trim={0 6cm  0 6cm},clip]{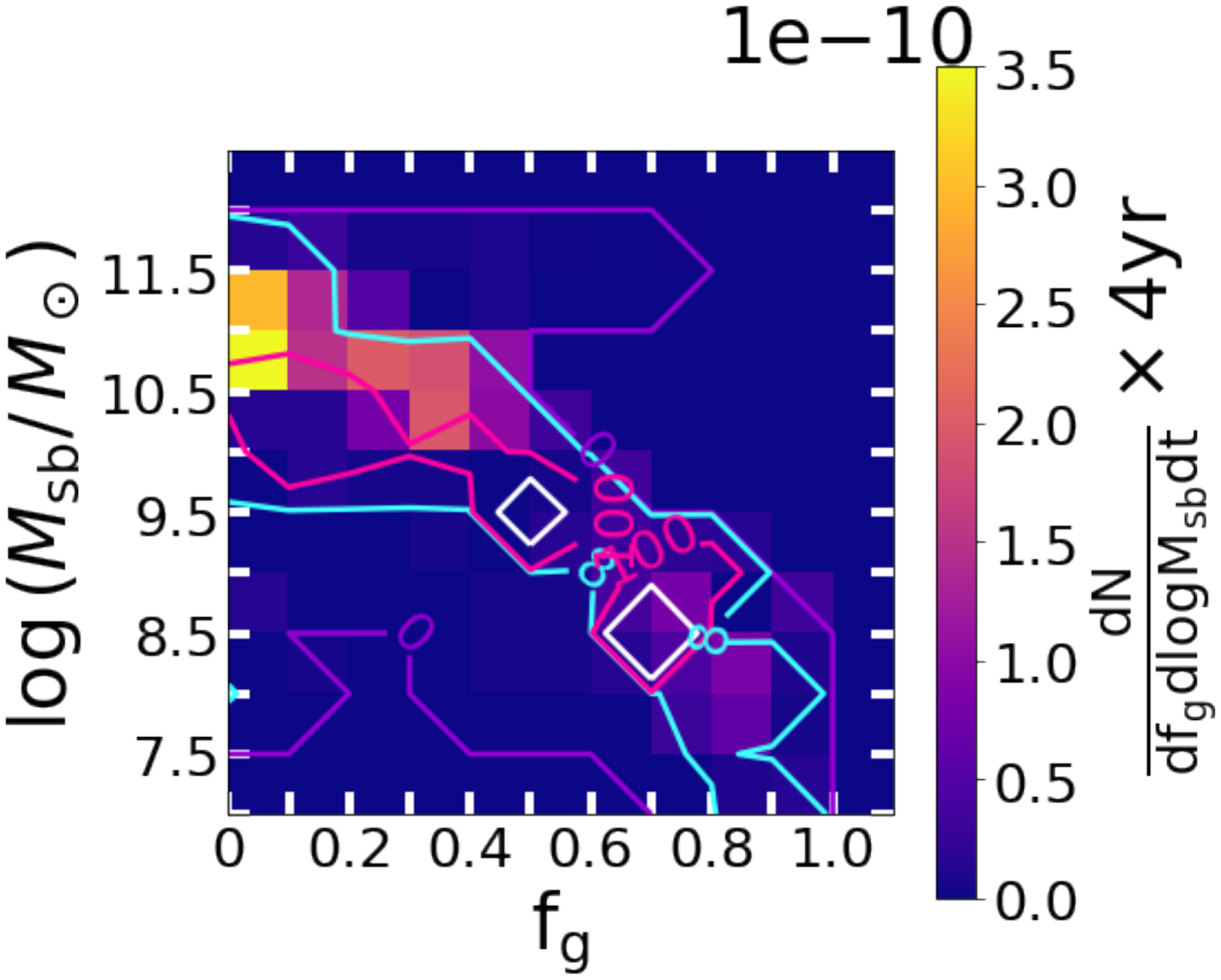}
            \includegraphics[width=0.49\textwidth , trim={0 6cm  0 6cm},clip]{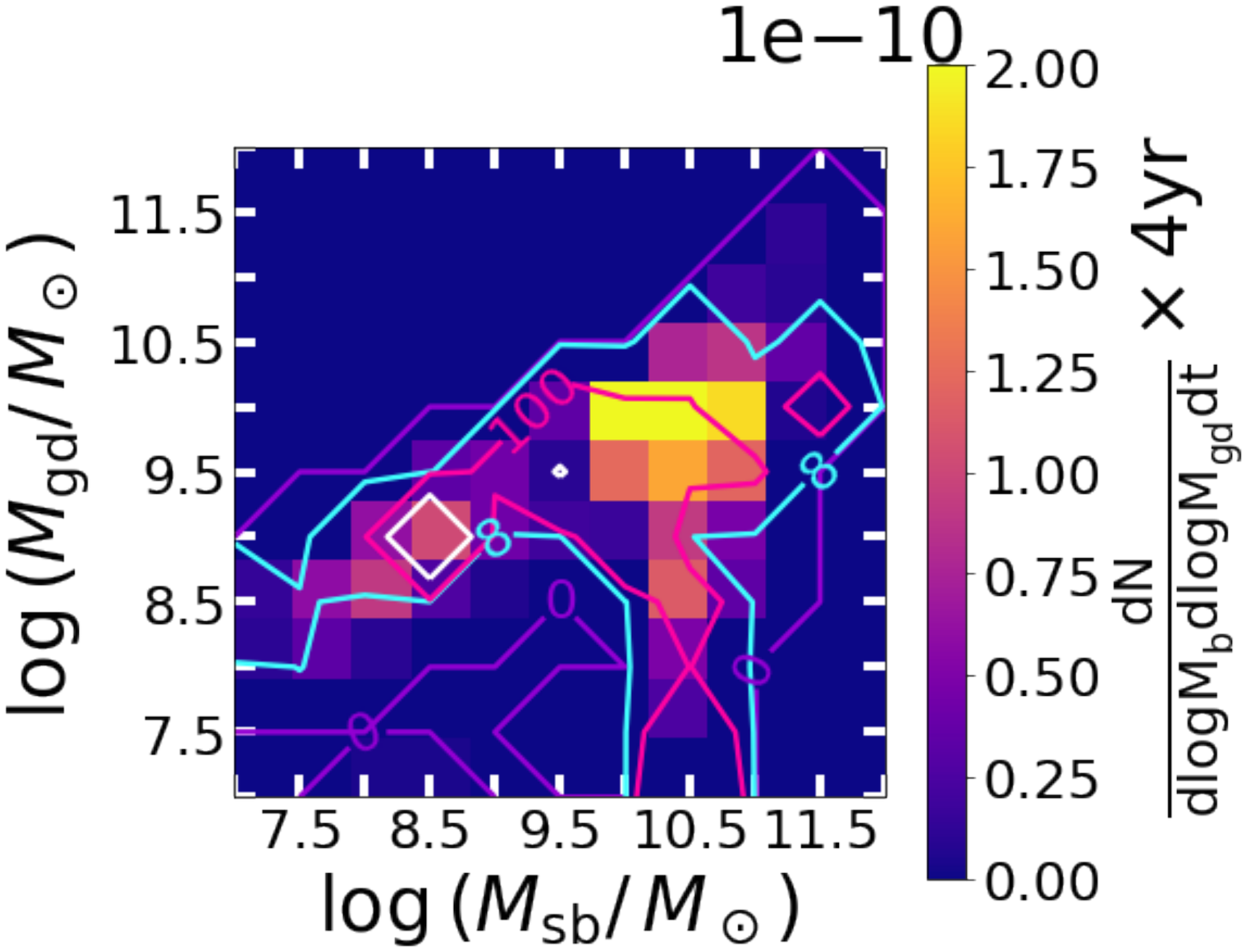}
\caption{As in Fig.~\ref{fig:2D_snr}, but now showing the results when
RF is included in the DF calculation. In this case, the MBHB
coalescences are most frequently found at low \fg\ and high
\Mtot\ (which corresponds to high \Msb). Therefore, most mergers will
 have low SNR and will be undetected by \textit{LISA}. These panels
 show the results for the prograde, $e_i < 0.2$ orbital
 configurations. The results are qualitatively similar for the remaining configurations.}
\label{fig:2D_snr_RF}
\end{figure*}

Interestingly, Fig.~\ref{fig:ratio_RF} shows that the nominal MBHB coalescence
fraction, calculated in the absence of RF, can in some cases be \textit{increased} in the presence of RF. This is the case for all data points with the value of the ratio of coalescence fraction $>1$ and happens 
when the \Ms\ is on a retrograde orbit and the merger galaxy has $f_g \la 0.3$ (panel
b). The enhanced coalescence fractions also correspond to more massive  \Mp s that tend to reside in galaxies with large bulge masses \citep[panels c and d;][]{Nelson2019, TNG50_a, TNG50_b} and in binaries with low $q$ (panel e).


This can be understood as follows. The RF (and negative DF) effects are usually not important for MBHs
on retrograde orbits because of the large relative velocity between the MBH and the galactic gas disk. If however the MBH is on a sufficiently eccentric orbit, then its relative velocity can fall into the appropriate range to satisfy the negative DF criteria (LBB20b; their Eq. 1). In the context of the systems considered here, the negative DF criteria are satisfied for lower mass sMBHs orbiting in lower gas density (and lower gas fraction) galaxies. In such systems negative DF accelerates the sMBH at pericenter, leading to a further increase of its orbital eccentricity (LBB20b). As a result, the sMBHs on these orbits
reach small pericentric distances and start to interact with the
circumbinary disk in the nucleus of the merger galaxy. This causes
their eccentricities to grow even more and ultimately, the \Ms\ to
plunge into the \Mp.  According to Fig.~\ref{fig:ratio_RF}, the
coalescence fraction of retrograde sMBHs in host galaxies with a large stellar bulge ($M_{\rm sb}> 10^{\rm  10} {\rm M_{\odot}}$) is boosted the most. This is because stellar DF helps to shorten the evolution time before the sMBHs enter the circumbinary disk and plunge into the pMBHs. Recall that galaxies with massive bulges also tend to have more massive pMBHs, thus explaining the dependence on this parameter in panel (c) and a preference for low mass ratio systems in panel (e).

The resulting integrated MBHB coalescence rates that include the
effects of RF are listed in the 4th column of
Table~\ref{tab:rate}. The largest rate is still found for MBHs moving
on prograde, low $e_i$ orbits, but it is now $dN_{\mathrm{mer}}/dt=0.1$~yr$^{-1}$, or $22\%$ of the value
calculated in the absence of RF. If the sMBHs are on prograde and high
$e_i$ orbits, the RF affected $dN_{\rm mer}/dt$ drops to
$0.04$~yr$^{-1}$, $13\%$ of its original value, which is the largest
reduction of the four configurations. Radiation feedback has the
smallest impact on the coalescence rate for systems with sMBHs on
retrograde, low $e_i$ orbits. In this case, $dN_{\rm
  mer}/dt=0.09$~yr$^{-1}$ which is only a 50\% reduction. MBHs on
these orbits are least affected by the RF due to the large relative
velocities of these orbits, which makes them largely immune to the effects
of RF. 

\subsection{The Effects of Radiation Feedback on the \textit{LISA} Detection Rates}
\label{sub:RF_LISA_rate}

We find that RF significantly reduces the rate of MBHB mergers
found in our TNG50-3 derived sample of post-merger galaxies. In
addition, the MBHs that do merge are now found
in low \fg\ galaxies. As mentioned in Sect.~\ref{sec:property_lisa},
there is an apparent anti-correlation between \fg\ and \Mtot\ in the TNG50-3
galaxies. Thus, a low \fg\ typically corresponds to galaxies with
larger \Mtot\ \citep{Nelson2019, TNG50_a, TNG50_b}, the mass end more challenging to detect with \textit{LISA} (see Figure~\ref{fig:dist_fg}). Therefore, we expect that the RF effects will most significantly reduce the merger rates of the lower mass MBHs, that are considered prime targets for \textit{LISA}.

These effects are illustrated in Figure~\ref{fig:2D_snr_RF} which
shows the two-dimensional differential number of mergers expected in
four years overplotted with contours of the \textit{LISA} SNR,
calculated as in Sect.~\ref{sec:property_lisa}. In the presence of RF the largest number of
mergers is expected for large \Mtot\, larger
$q$ and low \fg. The top two panels of this Figure show 
that the majority of mergers now fall outside the SNR$=8$ detection threshold, in contrast to the outcome of the calculation which does not include the impact of RF (see the corresponding
panels in Fig.~\ref{fig:2D_snr}). The bottom two panels show that any detection
\textit{LISA} can make will be tightly confined to high mass and
gas-poor host galaxies. Indeed, the effects of RF may make \textit{LISA}
detections in any gas-rich galaxy extremely rare.

The final column of Table~\ref{tab:rate} shows that the total predicted \textit{LISA}
detection rates from the RF calculations are $0.02$~yr$^{-1}$, independent of the orbital configuration. Thus, only
20\% of the prograde, low $e_i$ mergers would be detectable by
\textit{LISA}, but about half of the high $e_i$ mergers would be
detectable. Therefore, as expected, we find that there is a
significant reduction in the fraction of merging MBHBs that would be
detectable by \textit{LISA} if RF plays an important role in the
DF phase of the orbital decay.


\section{Discussion}
\label{sect:discuss}
\subsection{Comparison with Results in the Literature}
\label{sub:implications}

In previous sections we provide predictions for the \textit{LISA} detection rate and the properties of MBHBs that are most likely to become loud \textit{LISA} sources using data from the TNG50-3 simulation as input. TNG50-3 data provide a large parameter space of MBH and merger galaxy properties: pMBH masses in the range of $(10^{\rm 6}, 10^{\rm 9.5}) M_{\rm \odot}$, mass ratios $(10^{\rm -3}, 1)$, central gas number densities $(10^{\rm -3.5}, 10^{\rm 4.5})\, {\rm  cm^{-3}}$, and redshifts from 0 to 6. 



The \textit{LISA} detection rate predicted by our model ($\sim 0.3\, {\rm yr^{-1}}$ in the absence of RF, see Section\,5) is comparable to but somewhat lower than those found by \citet{Salcido2016} and \citet{Katz2019}. \citet{Katz2019} used results from the cosmological simulation Illustris, with  evolution times for MBH pairs with separations  $\lesssim$\,kpc calculated using semi-analytic models described by \citet{DA2017} and \citet{KBH2017}. 

The model by \citet{DA2017} treats the host galaxy as a singular isothermal sphere of stars and assumes that the sMBH remains embedded in a core of stars with a mass of $10^{\rm 3}$ times its own mass. The stellar remnant around the sMBH causes it to sink faster towards the galaxy’s center. This model takes into account the stellar DF, LC scattering, and the GW emission, but does not consider the effect of gaseous DF, radiative feedback, and the interaction with the circumbinary disk. 

The model by \citet{KBH2017} numerically integrates orbital evolution of MBH pairs from large separations until coalescence. In this model, the stellar DF is implemented following \citet{C1943}, and thus assumes that the perturber is moving in a uniform stellar environment and only stars that are moving slower than the perturber contribute to the DF. However, depending on the bulge properties, the stars that are moving faster than the sMBH can also contribute to the stellar DF \citep{AM2012}. After the DF stage, the LC scattering and the interaction with the circumbinary gas disk,  the GW emission continues to harden the binary until coalescence. This model does not take into account the stellar DF from fast moving stars, the gaseous DF, and the effect of radiative feedback.

Overall, Katz et. al. predict a \textit{LISA} detection rate of $\sim  0.5 - 1 {\rm yr^{-1}}$ for MBHBs with masses larger than $10^{\rm 5} M_{\rm \odot}$. Their prediction is slightly larger than ours, which is as expected, since the minimum mass in {\it Illustris} simulation is $\sim 10^{\rm 5}\, M_{\rm \odot}$. Thus, their work includes a broader range of binary mass, especially at the lower end, which makes their prediction for the \textit{LISA} detection rate higher.

 \citet{Salcido2016} use results from the cosmological hydrodynamic simulation suite EAGLE. They assume constant delay times between the galaxy merger and MBH coalescence, based on the gas content of the merger remnant galaxy. If the galaxy is gas-rich they assume a delay of 0.1 Gyr, or, if the galaxy is gas-poor, the delay is set to 5 Gyr. They also consider a variation of the evolution model where a prescription for the expected delays has been included after their host galaxies merge, and find that the merger rate is similar in all models. The MBH pair evolution time however depends on many more properties besides the gas fraction, hence, an assumption of constant evolution times makes the uncertainty in the {\it LISA} detection rate relatively large. \citet{Salcido2016} predict the  \textit{LISA} detection rate to be $\sim 2 \,{\rm yr^{-1}}$, six to seven times higher than ours, because their assumed time delays are on average shorter than the ones we calculate. Furthermore, the potential \textit{LISA} detections by \citet{Salcido2016} are mostly contributed by coalescences between seed mass black holes merging at redshifts between $1$ and $2$. This indicates that the detection rate they calculate is mainly for MBHs with masses less than $10^{\rm 6}\, M_{\rm \odot}$, which also explains their higher predicted detection rate.

More recently, \citet{Marta2020} used the cosmological simulations
  HORIZON-AGN \citep{Dubois2014b} and NEWHORIZON \citep{Dubois2021},
  to estimate the merger rate of MBHBs. Both simulations use the MBH
  seed mass of $\sim 10^{\rm 5} M_{\rm \odot}$, similar to TNG50, but
  represent different cosmological volume sizes that provide more
  complete census for either the high mass galaxies hosting
  $>10^7\,M_\odot$ MBHs (HORIZON-AGN) or the lower mass galaxies with
  $<10^7\,M_\odot$ MBHs (NEWHORIZON). Similar to this work, \citet{Marta2020} use
  a semi-analytic model to follow the evolution of MBH pairs below the
  spatial resolution of each simulation. For HORIZON-AGN, \citet{Marta2020} predict the cumulative MBHB coalescence rate measured by an observer at $z=0$ of $\sim  0.5$~yr$^{-1}$, based on a model most closely comparable to ours. This is similar to the rate calculated in this work, in the scenario when RF is not taken into account. The cumulative MBHB coalescence rate for NEWHORIZON is $\sim 1$~yr$^{-1}$, about two times larger than our prediction for MBH coalescence rate in the absence of RF.

 

\begin{figure}[t]
\centering
            \includegraphics[width=0.49\textwidth]{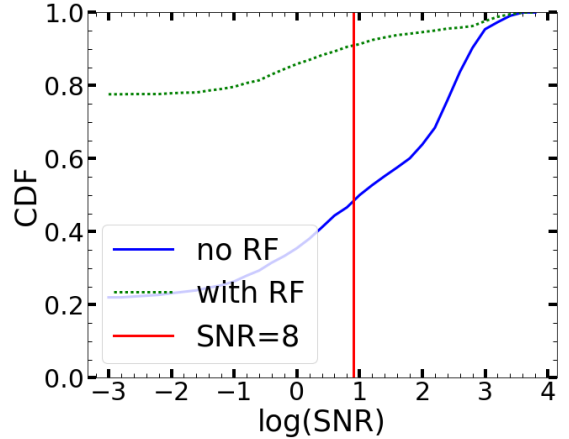}      
\caption{Cumulative distribution function (CDF) of the SNRs of all MBHB mergers considered in this calculation in the absence/ presence of RF. The vertical red line indicates the \textit{LISA} detection threshold used in this paper (${\rm SNR= 8}$). }
\label{fig:dist_SNR}
\end{figure}

From fully semi-analytic models, that rely on their own MBH seeding prescriptions instead on those of cosmological simulations,  \citet{K2016} and \citet{Berti2016} predict  \textit{LISA} detection rates $\sim 8$~yr$^{-1}$. This rate corresponds to the physical scenario where MBHs form from heavy seeds. We focus on this seeding prescription in our comparisons because it most closely corresponds to the MBH seeding prescription in TNG. The models in both mentioned studies include dynamical friction, stellar scattering, viscous drag, gravitational wave emission, and take into account a possibility of triple MBH systems. The model by \citet{K2016} is adopted from \citet{B2012}, which was subsequently improved by \citet{Sesana2014}, \citet{An2015a} and \citet{An2015b}. We are using the same gas disk and stellar bulge density profiles as \citet{K2016} but neglect the DF from the stellar disk, which we showed is negligible relative to the stellar bulge and gas disk (LBB20a). 


It is interesting to note that predictions for the \textit{LISA} detection rates are generally higher when calculated from models that use semi-analytic heavy MBH seed prescriptions as a starting point instead of cosmological simulations. This is caused by the differences in the MBH seeding approaches. Namely, the seeding in semi-analytic prescriptions is not limited by numerical resolution and they can generate arbitrarily small MBH seeds (usually $\sim 10^{\rm 4} M_{\rm \odot}$ in the heavy-seed scenarios). Therefore, semi-analytic prescriptions can populate with MBHs the low mass end of host galaxies, that are the major sources of coalescences at high redshift, where dwarf galaxies frequently merge into larger central galaxies. Cosmological simulations like EAGLE, Illustris and TNG are limited by their mass resolution and can only produce MBH seeds with mass $\gtrsim 10^{\rm 5} M_{\rm \odot}$ at redshifts smaller than those in semi-analytic prescriptions. As a result, the studies relying on cosmological simulations miss coalescences of low mass galaxies at high redshift, and this is the major reason for their comparatively low \textit{LISA} detection rates.

\subsection{Impact of Radiative Feedback on the LISA Detection Rate}
\label{sub:impact_RF_rate}

An aspect unique to this study is the calculation of the cosmological MBHB coalescence and \textit{LISA} detection rates when the effects of radiative feedback are taken into account. If negative gaseous DF operates in real galaxies as described in the literature so far \citep{PB2017,G2020,T2020}, this should result in longer inspiral times from $\sim $ kpc scales and reduction of the MBHB coalescence rate by $92\%$ and the \textit{LISA} detection rate to $0.02\, {\rm yr^{-1}}$. While the quoted rate worrisomely implies less than one detection in 4 years of nominal \textit{LISA} mission time, we emphasize that our model provides a conservative estimate of the LISA detection rates, due to the limited MBH mass range in TNG50-3. The relative reduction of the MBHB coalescence rate however is a robust prediction, as long as theoretical models capture how DF operates in the presence and absence of RF.

This is interesting in light of the finding discussed earlier, that to the first order, the DF alone is sufficient to determine the distribution of total evolution time for a population of MBH pairs (as illustrated by Figure~\ref{fig:dist_tbound}). In other words, for a population of MBH pairs, the approximate distribution for $t_{\rm tot}$ can be recovered even if the evolution due to the loss cone scattering, viscous drag and gravitational wave emission is neglected. Because $t_{\rm tot}$ is closely related to the rate of the MBHB coalescences, the measurement of the latter should provide direct insights in how DF operates in mergers. This is important because, although widely embraced as a mechanism for orbital evolution of MBH pairs, DF is still a theoretical concept and is yet to be verified in observations.

For example, low MBHB coalescence rates inferred from \textit{LISA} measurements would necessitate consideration of the effects of radiative feedback. As shown in Figure~\ref{fig:dist_SNR}, in the absence of RF nearly $60\%$ of all MBHBs from TNG50-3 have $SNR>8$. Alternatively, in the presence of RF only $10\%$ of events can be detected by \textit{LISA}. 
Intriguingly, we also find that in the presence of RF nearly the same percentage ($\sim 25 \%$) of \textit{LISA} detections correspond to each of the four orbital configurations in our study, if all are equally represented  (see Section\,~6). This means that if the importance of RF is established for the observed GW events, it should also be possible to recover the underlying distribution of orbital eccentricities, given that in that case orbits of all eccentricities are equally likely to result in the coalescence and detection.

\subsection{Impact of Simplifying Assumptions}
\label{sub:assumptions}

The advantage of our semi-analytic model is the ability to run
simulations quickly over a wide range of galaxy and MBH orbital properties at
the cost of making some simplifying assumptions. The potential impact
of our assumptions on the dynamical aspects of MBHB evolution is discussed in our previous works (LBB20a,b). In this section, we consider the possible effects of these assumptions on the \textit{LISA} detection rates.

In this work, we assume the \Mp\ is fixed at the center of the host galaxy. If the  motion of the \Mp\ and its orbital decay due to DF forces are included in the simulations, the resulting \tevol\ would be shorter, consequently increasing the coalescence rate and \textit{LISA} detection rate. This effect would be strongest in comparable mass MBH pairs and weaker in those with small $q$. Allowing the \Mp\ to move around the center of mass would also increase the number of high redshift \textit{LISA} detections since MBHBs at high redshifts tend to have larger mass ratios. 

We furthermore posit that the orbit of the \Ms\ is always in the midplane of the galaxy. For the \Ms\ on an inclined orbit, that takes it outside of the gas disk, \tevol\ generally increases since the gaseous DF force is less efficient. This results in lower coalescence and \textit{LISA} detection rates independent of the presence of RF.


In our simulation, we assume there is no stellar remnant around the sMBH. The pertinent question is what fraction of the remnant nuclear star cluster is still bound to the sMBH when it reaches the starting point for our simulations (a separation of $\sim 1 $ kpc). The answer to this question differs, depending on the specific study and model used. For example, this question was examined by \citet{KBH2017}, who assumed that the mass of the sMBH and the remnant cluster decreases as a power law. They find that after a dynamical time, the stellar enhancement runs out and the hardening rate approaches that of a bare sMBH by $\sim 1$ kpc. In another model \citep{DA2017}, the mass of the remnant stellar cluster around the sMBH can be significant at $\sim 1 $ kpc separation. If so, the orbital evolution time of the sMBH would be shorter than calculated in our work.

Another assumption adopted in this work is that MBHs do not grow in mass during their orbital evolution from kpc scales toward coalescence. The mass ratios of MBH pairs in the calculations change, on average, by $\leq \sim 60 \%$ during the DF dominated stage \citep{LBB21a}. However, the change in mass ratio during the evolution in the circumbinary disk depends sensitively on the binary accretion model. Many simulations have shown that the accretion occurs preferentially onto the smaller MBH in a binary because being closer to the inner rim of the circumbinary disk allows it to more easily capture gas \citep{Art1994, Gun2002, Hayasaki2007, Roedig2011, Nixon2011, Farris2014}. This phenomenon speeds up orbital decay in the disk as the sMBH may grow at a higher rate and the mass ratio of the binary may become larger \citep{Siwek2020}. By assuming a fixed mass ratio throughout the evolution, we effectively provide a more conservative coalescence rate and the \textit{LISA} detection rate. If the accretion driven mass growth of MBHs by accretion is taken into account, we expect both rates to increase.

We assume that the DF does not contribute to the orbital decay when the separation between the two MBHs is smaller than the influence radius. This potentially leads us to underestimate the rate of orbital inspiral, since DF (and especially the stellar DF exerted by the bulge on the sMBH) can be of the same importance as the the loss cone scattering or viscous drag at separations where one mechanism transitions to another. If instead we took into account the contribution of DF to the orbital decay below the influence radius, the coalescence rate and the \textit{LISA} detection rate would be higher. We expect this to be the case regardless of the presence or absence of the effects of radiative feedback. 

We do not account for the DF generated by the stellar disk in the merger remnant galaxy, since in our previous study we found it to be negligible in most cases \citep{LBB20a}. \citet{Bonetti2021} have shown however that in some situations (i.e., when the gas fraction is low) the DF from the stellar disk could be important. Thus, taking the DF from stellar disks into account would result in shorter time to coalescence for MBH pairs in such galaxies and would consequently boost the coalescence and \textit{LISA} detection rates. This is especially true for systems affected by radiative feedback, in which the DF from stellar disks could counter the negative DF from the gas.

We neglect the dynamical evolution and coalescences that arise as a consequence of the formation of MBH triplets. At high redshift, where the merger rate of galaxies is higher, it is possible for a third MBH to inspiral and join the orbital decay of a MBHB. The system may undergo the Kozai-Lidov oscillations, which could boost the eccentricity of the central binary and in such way speed up its coalescence \citep{Kozai1962}. Besides the Kozai-Lidov oscillations, the chaotic three-body interactions can also boost the coalescence rate \citep{Bla2002,Hoffman2007, Amaro2010, Kul2012, Bonetti2016, Ryu2018}. Consequently, the three-body interactions can bring a sizeable number of stalled MBHBs to coalescence when other mechanisms fail to work \citep{Bonetti2018, Bonetti2019}. According to \citet{Bonetti2019}, the interaction between MBH triplets can increase the coalescence rate by $\sim 40\%$ in the case of the heavy-seed model. 

The description of RF used here is based on idealized simulations that assume an isolated MBH moving on a linear trajectory in a uniform gas density and an infinite background medium \citep{Err2019, T2020}. It is possible that the non-uniform environment in the aftermath of the galactic merger will perturb the smooth orbital decay of the sMBH \citep{Fiacconi2013} and weaken the effects of RF, in particular the acceleration due to negative DF. Similarly, when the radius of the H\textsc{II} region around the sMBH exceeds the half thickness of the gas disk, radiation can escape the disk plane, reducing the impact of RF. However, three-dimensional simulations of MBH evolution that include both gaseous DF and RF also show a weakening of the DF force \citep{Sijacki2011, Souza2017}. Thus, current work indicates that relaxing the assumptions in the RF model will reduce but not erase the negative DF effects of RF. We conclude that future studies of MBH coalescence rates should carefully consider the potential impact of RF during the DF decay phase of the sMBH.
 
In this work, we considered the effect of RF on DF only, but studies have shown that the RF can also affect the dynamical evolution of sMBHs in the circumbinary disks. According to \citet{RF1pc}, when there is no tidal cavity in the disk, the sMBHs can accrete at high rates. The resulting strong winds collide against the disk, pushing the gas away from the binary which stalls the binary migration. On the other hand, when there is already a gap in the disk opened by the sMBH, the RF does not affect the evolution of the binary and the structure of the disk since the wind launched from the sMBH can escape perpendicularly to the disk though the tidal cavity without further disturbing the disk \citep{RF1pc}. Our model assumes the gap-opening regime, thus taking into account the RF in the circumbinary disks should not affect our results.

The TNG simulation, and consequently our study, do not capture MBHs with mass lower than about $10^{\rm 6}\,M_{\rm \odot}$. This leads to an underestimate of the coalescence and detection rates, particularly on the low mass end. This seems to be a limitation shared by multiple models that use cosmological simulations as their input, as discussed in Section~\ref{sub:implications}. It is worth noting that our rates are in broad agreement with other such models \citep{Salcido2016, Katz2019}, even if our detailed approaches differ.

Overall, we expect that our model provides a conservative estimate for the MBHB coalescence rate and the \textit{LISA} detection rate, and we expect these values to be higher if the above assumptions are relaxed.

\section{Conclusions}
\label{sec:concl}




In this work we evaluate the cosmological coalescence rate and the detection rate of MBHBs targeted by the \textit{LISA} GW observatory. Our calculation starts with a population of initially gravitationally unbound MBH pairs, with separation of about 1\,kpc, drawn from the TNG50-3 cosmological simulation. We then follow their dynamical evolution all the way to coalescence under the influence of the stellar and gaseous dynamical friction and at smaller separations -- stellar scattering, viscous drag from the circumbinary disk and the GW emission. We also explore the effects of radiative feedback from the accreting MBHs on their coalescence and detection rates. The main results of this study are summarized below.

\begin{enumerate}
\item  We find that for a majority of modeled MBH pairs DF determines the total evolution time (see Figure~\ref{fig:dist_tbound}), and hence their cosmological coalescence rate, whereas the impact of the physical mechanisms that operate at smaller orbital separations is small. This means that the MBHB coalescence rate, obtained from the GW measurements, will first and foremost provide {\it statistical} constraints on the efficiency of DF in merger galaxies. This is an important opportunity to verify the importance of DF in observations which, although widely embraced as a mechanism for orbital evolution of MBH pairs, is still a theoretical concept. 


\item Based on our models that do not account for the MBH radiative feedback we find that the MBHB coalescence rate is $\lesssim 0.45$~yr$^{-1}$, and the \textit{LISA} detection rate is $\lesssim 0.34$~yr$^{-1}$. Most \textit{LISA} detections should 
have a characteristic SNR $\sim 100$ and originate from galaxies at redshifts $1.6 \leq z \leq 2.4$. They will correspond to binaries with masses $10^{\rm 6} - 10^{\rm 6.8}\,M_{\rm \odot}$ and comparable mass ratios,  $q = 0.5 - 1$, located in gas-rich galaxies with gas fractions in the range $0.6-0.9$. In this case, high gas fractions bode well for a chance to detect the  associated electromagnetic counterparts.

\item We find a significant reduction in the number of merging MBHBs if RF plays an important role in the DF phase of the orbital decay. In the presence of RF, the MBHB coalescence rate is reduced by $78\%$ (to $\lesssim 0.1$~yr$^{-1}$), and the \textit{LISA} detection rate is reduced by $94\%$ (to $0.02$~yr$^{-1}$). In this case, we expect most \textit{LISA} detections to have a lower characteristic SNR$ \sim 10$ and to originate from galaxies at a comparatively lower redshift, $1\leq z \leq 2$. They will correspond to binaries with high masses, $10^{\rm 7} - 10^{\rm 9} M_{\rm \odot}$, and $q = 0.5 - 1$, located in relatively gas-poor galaxies with gas fractions $0 - 0.1$. Combined with the low GW detection rates, such low gas fractions could make the detection of the associated electromagnetic counterparts difficult, making this observationally the most challenging scenario.

\item  In the absence of RF, the MBHB systems that become \textit{LISA} sources are more likely to evolve in prograde orbital configurations, with either low or high eccentricity. For example, if prograde and retrograde configurations were initially equally represented, we expect $\sim 60\%$ of \textit{LISA} detections to come from MBHBs in prograde orbits. In the  presence of RF, we expect equal fractions ($\sim 25 \%$) of \textit{LISA} detections to come from MBHBs in prograde and retrograde, low or high initial eccentricity orbits, if all were equally represented initially.  Thus, if the importance of RF is established for the observed GW events, it should be possible to recover the underlying distribution of orbital orientations and eccentricities.

\end{enumerate}

We emphasize that our model provides a conservative estimate of the \textit{LISA} detection rates, due to the limited MBH mass range in TNG50-3. In the case when the effects of RF are not taken into account, our predicted rates are comparable to the models in the literature that draw their MBH pairs from cosmological simulations. Thus, predictions from this class of models seem to be relatively robust against the differences in cosmological simulations and individual model assumptions. The striking reduction of the MBHB coalescence rate in the presence of RF is a prediction unique to this work, without a readily available comparison in the literature. We expect it to be robust, as long as current theoretical models capture the salient properties of DF in the presence of RF.

Further advances in understanding the efficiency of DF on MBHs in merger galaxies, and thus their coalescence rate, can also be made with electromagnetic observations. The current and future X-ray (e.g., \textit{eROSITA}, \citealt{erosita}; \textit{Athena}, \citealt{athena}), radio (e.g., ngVLA, \citealt{ngvla}; SKA; \citealt{ska}), and optical/IR observatories (e.g., \textit{JWST}, \citealt{jwst}) will dramatically increase the population of known dual AGNs. Dual AGNs at separations $\lesssim$\,kpc in particular can provide a test of DF models and timescales, where an overabundance of systems on these scales would be consistent with expectations for the negative DF. On the other hand, some of these systems are progenitors of the \textit{LISA} sources that will merge within a Hubble time and can therefore be used to predict the \textit{LISA} detection rate of MBH mergers. In order to do so, one needs to know what types of merger remnant galaxies are most likely hosts to electromagnetically bright dual AGNs and later on, MBH coalescences detectable by \textit{LISA}. We defer this investigation to our upcoming publication.

\begin{acknowledgments}
T.B. acknowledges the support by the National Aeronautics and Space
Administration (NASA) under award No. 80NSSC19K0319 and by the
National Science Foundation (NSF) under award No. 1908042. 
\end{acknowledgments}

\appendix

\section{Gravitational Wave Emission and Detection}
\label{app:GW_emission}

\subsection{\textit{LISA} Sensitivity Curve}

We use the sky-averaged \textit{LISA} sensitivity curve presented in
the ``\textit{LISA} Strain Curves" document (LISA-LCST-SGS-TN-001),

\begin{equation*}
\label{eq:sn}
S_n(f)=\frac{10}{3L^{\rm 2}}\left ( P_{\rm OMS} + \frac{4P_{\rm acc}}{(2\pi f)^{\rm 4}}\right )\left ( 1+\frac{6}{10}\left ( \frac{f}{f_{\rm \star}}\right )^{\rm 2} \right ) + S_{c} (f),
\end{equation*}
%
%
where the \textit{LISA} arm-length is $L=2.5\times 10^{\rm 9}$~m, and $f_{\rm \star}=19.09$ mHz, with
single-link optical metrology noise
\begin{equation}
\label{eq:poms}
P_{\rm OMS}=2.25\times 10^{\rm -22}\left [ 1+\left( \frac{2\,{\rm mHz}}{f}\right )^{\rm 4}\right] {\rm m}^{\rm 2} {\rm Hz}^{\rm -1},
\end{equation}
and single test mass acceleration noise

\begin{equation}
\label{eq:pacc2}
P_{\rm acc}=9\times 10^{\rm -30}\left [ 1+\left( \frac{0.4\,{\rm mHz}}{f}\right )^{\rm 2}\right] 
 \left[ 1+\left( \frac{f}{8{\rm mHz}} \right )^{\rm 4}\right] {\rm m}^{\rm 2} {\rm Hz}^{\rm 3}.
\end{equation}
Furthermore, we account for the average confusion noise from galactic compact binaries,  $S_{c} (f)$ \citep{Cornish2018},
\begin{equation}
\label{eq:sc}
S_c (f)=Af^{\rm -7/3}e^{f^{\rm \alpha}+\beta f {\rm sin}({\rm \kappa}
  f) }[1+{\rm tanh}({\rm \gamma}(f_{k}-f))]\ {\rm Hz}^{\rm -1},
\end{equation}
where $A=9\times 10^{\rm -45}$, $\alpha=0.138$, $\beta=-221$, $\kappa=521$, $\gamma=1680$, and $f_{k}=0.00113$~Hz in a four year mission time.

\subsection{Calculation of SNR}

We calculate the SNR accumulated during the inspiral stage of MBHBs following \citet{Barack2004}
\begin{equation}
\label{eq:snr}
({\rm SNR})^{\rm 2}=\sum_{n}\int \frac{h_{c,n}^{\rm 2}}{f_{n}S_{n}(f_{n})}{\rm d}{\rm ln} f_{n},
\end{equation}
where $f_n=nf_{orb}$ is the frequency of the $n$-th harmonic in the GW
spectrum, and $S_{n}(f_{n})$ is the power spectral density of
\textit{LISA} averaged over sky-inclination-polarization. The
characteristic strain of the $n$-th harmonic
\begin{equation}
\label{eq:hcn}
h_{c,n}=\frac{1}{\pi d_{L}}\sqrt{\frac{2G\Dot{E}_{n}}{c^{\rm 3}\Dot{f_{n}}}},
\end{equation}
where $d_{L}$ is the luminosity distance to the GW source, and $\Dot{E}_n$ is the amount of power in the $n$-th harmonic given by \citet{Peters1963}

\begin{equation}
\label{eq:en}
\Dot{E}_n=\frac{32G^{\rm 7/3}}{5c^{\rm 5}}(2\pi M_{\rm chirp}f_{\rm orb})^{\rm 10/3}g_n(e),
\end{equation}
and 
\begin{equation*}
\label{eq:gn1}
g_n(e)=\frac{n^{\rm 4}}{32}\big [ \big (J_{n-2}(ne)-2eJ_{n-1}(ne)+\frac{2}{n}J_{n}(ne)+2eJ_{n+1}(ne)-J_{n+2}(ne)\big )^{\rm 2}
\end{equation*}

\begin{equation}
\label{eq:gn4}
+(1-e^{\rm 2})\big (J_{n-2}(ne)-2J_{n}(ne)+J_{n+2}(ne)\big )^{\rm 2}+\frac{4}{3n^{\rm 2}}J^{\rm 2}_{n}(ne)\big ].
\end{equation}

Here $J_n$ is the $n$-th order Bessel function of the first kind.

\end{document}